\documentclass[fleqn,usenatbib,twocolumn]{mnras}

\usepackage{newtxtext,newtxmath}

\usepackage[T1]{fontenc}

\DeclareRobustCommand{\VAN}[3]{#2}
\let\VANthebibliography\thebibliography
\def\thebibliography{\DeclareRobustCommand{\VAN}[3]{##3}\VANthebibliography}

\usepackage{graphicx}	
\usepackage{amsmath}	
\usepackage{bm}
\usepackage{tabularx}
\usepackage{pdflscape}
\usepackage{color}
\usepackage{times}
\usepackage{hyperref}
\usepackage{colortbl}

\definecolor{Gray}{gray}{0.9}

\makeatletter
\newcommand{\HI}{{\rm H\,{\scriptstyle I}}}

\newcommand{\HeII}{{\rm He\,{\scriptstyle II}}}

\newcommand{\OII}{{\rm O\,{\scriptstyle II}}}
\newcommand{\OIII}{{\rm O\,{\scriptstyle III}}}
\newcommand{\OVI}{{\rm O\,{\scriptstyle VI}}}

\newcommand{\Rmnum}[1]{\expandafter\@slowromancap\romannumeral #1@}

\newcommand{\xHI}{x_{\mbox{\tiny H\Rmnum{1}}}}

\newcommand{\sigmaHI}{\sigma_{\mbox{\tiny H\Rmnum{1}}}}
\newcommand{\Muv}{M_{\mbox{\tiny UV}}}

\newcommand{\A}{\mbox{\AA}}

\title[Photometric light-echo tomography]{Photometric IGM Tomography: Efficiently Mapping Quasar Light Echoes with Deep Narrow Band Imaging}

\author[K. Kakiichi et al.]{
Koki Kakiichi,$^{1}$\thanks{E-mail: kakiichi@ucsb.edu (KK)}
Tobias Schmidt,$^{2}$ and
Joseph Hennawi$^{1,3}$
\\
$^{1}$Department of Physics, Broida Hall, University of California, Santa Barbara Santa Barbara, CA 93106-9530, USA\\
$^{2}$Observatoire Astronomique de l'Universit\'e de Gen\`eve, Chemin des Maillettes 51, Sauverny, CH-1290, Switzerland\\
$^{3}$Leiden Observatory, Leiden University, PO Box 9513, 2300 RA Leiden, The Netherlands
}

\date{Accepted XXX. Received YYY; in original form ZZZ}

\pubyear{2022}

\begin{document}
\label{firstpage}
\pagerange{\pageref{firstpage}--\pageref{lastpage}}
\maketitle

\begin{abstract}

In the standard picture, episodes of luminous quasar activity are directly related to supermassive black hole (SMBH) growth. The ionising radiation emitted over a quasar's lifetime alters the ionisation state of the surrounding intergalactic medium (IGM), enhancing the Ly$\alpha$ forest transmission -- so-called proximity effect -- which can be observed in absorption spectra of background sources. Owing to the finite speed of light, the transverse direction of the proximity effect is sensitive to the quasar's radiative history, resulting in `light echoes' that encode the growth history of the SMBH on Myr-timescales. In this paper, we introduce a new technique to photometrically map this quasar light echoes using Ly$\alpha$ forest tomography by using a carefully selected pair of narrow-band filters. A foreground narrow-band filter is used to measure Ly$\alpha$ forest transmission along background galaxies selected as Ly$\alpha$ emitters by a background narrow-band filter. This novel double narrow-band tomographic technique utilises the higher throughput and wider field of view of imaging over spectroscopy to efficiently reconstruct a two-dimensional map of Ly$\alpha$ forest transmission around a quasar. We present a fully Bayesian framework to measure the luminous quasar lifetime of a SMBH from photometric IGM tomography, and examine the observational requirements. This new technique provides an efficient strategy to map a large area of the sky with a modest observing time and to identify interesting regions to be examined by further deep 3D follow-up spectroscopic Ly$\alpha$ forest tomography.

\end{abstract}




\section{Introduction}\label{sec:intro}
Understanding the origin of supermassive black holes (SMBHs) is a long-standing problem in observational cosmology \citep[e.g.][]{Rees1978}. Recent wide field imaging surveys have discovered a dozen of $z\gtrsim7$ quasars \citep{Mortlock2011,Banados2018,Matsuoka2019,Yang2020,Wang2021}, which suggests that there is a too short time available to assemble enough SMBH mass of $\gtrsim10^9\rm M_\odot$ if we assume the Eddington-limited growth of a stellar-mass black hole from a massive stellar remnant. To explain their existence, theorists postulated the formation of massive seeds from the direct collapse black hole (DCBH) of a supermassive star or dense cluster of Pop III stars \citep[e.g.][for reviews]{Woods2019,Inayoshi2020}, or rapid black hole growth with super-Eddington accretion onto stellar-mass seeds \citep{Madau2014}. While the direct search of such enigmatic massive seeds may become possible with future wide-field transients and imaging surveys \citep{Whalen2013, Chen2014, Moriya2021} as well as with the future generations of gravitational wave facilities \citep{Shibata2016,Hartwig2016,Hartwig2018}, recent radiation magnetohydrodynamic simulations of accretion disks \citep{McKinney2014,Jiang2014,Jiang2019} suggest that a super-Eddington accretion is equally possible provided that there is enough material is being fed into the circumnuclear environment around the black hole \citep{Angles-Alcazar2021,Toyouchi2021,Inayoshi2021}. Observationally, \citet{Davies2019} argue evidence for the low radiative efficiency in $z>7$ quasars, consistent with the super-Eddington growth of the early SMBHs. In order to test the various formation scenarios, we ought to both observationally (1) test the existence of massive seeds and (2) to constrain the growth mechanism of the SMBHs.

Luminous quasar activity is closely linked to the growth history of SMBHs, which are believed to be powered by the gas accretion onto a central black hole. A quasar bolometric lightcurve $L(t)$ is related to the gas accretion rate $\dot{M}$,
\begin{equation}
L(t)=\epsilon\dot{M}c^2,
\end{equation}
where $\epsilon$ is the radiative efficiency which is $\epsilon\sim0.10$ for the standard thin accretion disc theory in general relativity \citep{Shakura1973,Novikov1973}. Comparison of the local total mass density of SMBHs and the total cosmic luminosity of quasars integrated over the age of the Universe, the so-called So\l{}tan argument \citep{Soltan1982,Kulkarni2019,Shen2020}, indicates that the most of the local SMBH mass is acquired during the luminous quasar phases for an average radiative efficiency of $\epsilon\sim0.08$ \citep[e.g.][]{Yu2002,Shankar2004,Ueda2014}. This makes the quasar lightcurve an excellent observational tool to test how SMBHs acquired their masses over their growth history.

The key characteristic timescale for the growth of a SMBH is the Salpeter timescale,
\begin{equation}
t_{\rm sal}=4.5\times10^7\left(\frac{\epsilon/(1-\epsilon)}{0.1}\right)\left(\frac{L}{L_{\rm Edd}}\right)^{-1}\rm\,yr,
\end{equation}
where $L_{\rm Edd}$ is the Eddington luminosity, which is equivalent to the $e$-folding time of the black hole mass growth $M_{\rm BH}=M_{\rm seed}\exp(t/t_{\rm sal})$. The time required for a stellar mass seed of $\sim100\,\rm M_\odot$ to grow $\sim10^9\rm\,M_\odot$ SMBH is approximately 16 e-foldings, $t\sim16 t_{\rm sal}\approx 7\times10^8\rm\,yr$. If quasars represent the major growth phase of SMBHs as suggested by the So\l{}tan argument, the quasar lifetime -- defined as the duration over which a quasar is active -- should be comparable to the Salpeter timescale. Estimates of the quasar lifetime range between $10^4$ and $10^8$ yr \citep[e.g.][]{Martini2004}. The recent observations of the line-of-sight proximity zone sizes of $z\sim3-6$ quasars indicate a short quasar lifetime of $t_{\rm age}\sim10^6\rm\,yr$ on average \citep{Morey2021,Khrykin2021} and $\sim1-10\%$ of the population shows even shorter lifetime of $t_{\rm age}\sim10^{4-5}\rm\,yr$ \citep{Eilers2017,Eilers2020,Eilers2021}. This calls the standard picture of SMBH growth into questions, meaning that the SMBH mass is too massive to be explained by the gas accretion during the lifetime of the quasar with Eddington-limited growth. However, the line-of-sight proximity effect is only sensitive to the most recent quasar activity. The fast relaxation time of a highly ionzied IGM by a quasar radiation to the ionized fraction ($\xHI\sim10^{-5}$) of the mean IGM is short, $t_{\rm relax}\simeq \xHI t_{\rm rec}\sim10^{4-5}\rm yr$, comparable to the equilibration\footnote{The equilibration timescale for $\HeII$ is longer, meaning that the line-of-sight $\HeII$ proximity zone can tolerate the quasar-inactive phase of $t_{\rm relax}\sim10^7\rm\,yr$.} timescale $t\sim\Gamma_{\rm HI}^{-1}$ (\citealt{Davies2020} see also \citealt{Khrykin2017}). This means that the line-of-sight proximity effect can only probe the duration of the most recent quasar activity if the quasar-inactive phase is longer than $t_{\rm relax}\sim 10^{4-5}\rm yr$.

Furthermore, using a statistical argument \citep[e.g.][]{Shen2009,White2012,Eftekharzadeh2015,Laurent2017,He2018,Timlin2018}, quasar clustering can constrain the average integrated quasar lifetime over the Hubble time, which is also referred to as the duty cycle. These observations suggest that the integrated quasar lifetime is approximately $\sim10^{7-8}\rm\,yr$, broadly consistent with the time required to grow SMBHs through quasar activities. As well, the statistical measurement of the transverse proximity effect in $\HeII$ Ly$\alpha$ forest by $z\sim2-3$ quasars -- enhanced $\HeII$ Ly$\alpha$ forest transmission in a background sightline by the ionization of a foreground quasar in the transverse direction -- provides a purely geometrical lower limit on the quasar lifetime of $t_{\rm age}>25\,\rm Myr$ \citep{Schmidt2017,Schmidt2018}, suggesting that the active phase of a quasar may be long enough to acquire sufficient mass through luminous mass accretion.  While these constraints are still weak, in order to reconcile both the line-of-sight/transverse proximitty effects and clustering measurements, variable quasar lightcurve of a SMBH is required. While many simulations indicate such episodic quasar phases are common owing to the intermittent gas accretion and the intense radiative and kinetic feedback from the quasar on the scale of host-galaxy and cosmological environment \citep[e.g.][]{Ciotti1997,Hopkins2008a,Novak2011}, the direct observational evidence still remains elusive.

The variable quasar lightcurve has a distinct impact on the ionization state of the circum- and inter-galatic medium (CGM and IGM) since the immense ionizing radiation from a quasar outshines the host galaxy and its surroundings. Because the speed of light is finite, the ionization state of the gas at a distance $r$ from the SMBH is sensitive to the ionizing output of the quasar activity at time $t=r/c$ in the past. This means the map of the ionization state of the gas at various distances from the SMBH can record {\it light echoes}, tracing directly the activity of quasar over the past history of the host galaxy. For example, \citet{Lintott2009} discovered emission from quasar-excited highly ionzied gas at the circum-galactic distance $\gtrsim10\,\rm kpc$,  called Hanny's Voorwerp, from a nearby spiral galaxy, arguing for the recent fading of quasar within $10^5$ years. The further observations of quasar light echoes in circum-galactic emission indicates the recent fading of quasar activity on $\sim10^5\rm yr$ timescale occurred in some nearby galaxies \citep{Keel2012,Keel2015,Keel2017}. A similar argument was made by \citet{Oppenheimer2018} who instead used the $\OVI$ absorbers in the CGM as an indicator of quasar-ionized gas around $z\sim0.2$ galaxies to demonstrate that variable quasar activities in past $\lesssim10^6\rm\,yr$ could explain their abundance. Conversely, the ubiquity of quasar-powered Ly$\alpha$ nebulae around $z\sim2-3$ quasars extending to $r\sim50\rm\,kpc$ argues for lifetimes of $>10^5$ yr \citep{Hennawi2007,Borisova2016,Arrigoni-Battaia2016} with the largest such $\sim~500\rm\,kpc$ nebulae \citep{Cantalupo2014,Hennawi2015} corresponding to $10^6$ yr. \citet{Hennawi2007} came to similar conclusions based on the anisotropic clustering pattern of optically thick \ion{H}{1} absorbers around quasars. Searching for the light echoes in the CGM can only provide a short baseline of $t\sim10-500{\rm\,kpc}/c\sim10^{4-6}\rm\, yr$, which is too short to probe the full quasar lightcurve on the scale of the Salpeter timescale. Furthermore, and perhaps most importantly, these CGM constraints suffer from ({\it i}) a degeneracy with the quasar opening angle (we know quasars are obscured in some directions) and ({\it ii}) we do not have a first principle model to robustly predict the physical state of CGM. This provides a strong motivation to focus searching for light echoes in the IGM where the physical state of the gas can be predicted ab initio from from cosmological simulations.

In order to probe the full quasar lightcurve comparable to the Salpeter timescale, we need to search for quasar light echoes on the scales of $\sim 1-10\rm\,Mpc$ around the host galaxy. The tomographic mapping of the IGM around a quasar using Ly$\alpha$ forest absorption along background sources provides a required technique to constrain the full quasar-active growth history of a SMBH. \citet{Adelberger2004} proposed to constrain the radiative history of quasar activities through the transverse proximity effect around a quasar using Ly$\alpha$ forest absorption along background galaxies \citep[see also][]{Visbal2008,Schmidt2019}. The IGM Ly$\alpha$ forest tomography \citep{Lee2014a,Lee2014b,Lee2018,Newman2020,Ravoux2020} makes it possible to map the impact of quasar's light echoes on the ionization state of the IGM \citep{Schmidt2019,Mukae2020a,Mukae2020b}, enabling us to probe the lightcurve on timescales of $10^6$ to $10^8$ yr because of the large Mpc-scale separation between the IGM and the central SMBH.

The method of mapping the quasar light echoes using the deep spectra of background galaxies via Ly$\alpha$ forest tomography is observationally expensive and time consuming, requiring a dedicated spectroscopic follow-up campaign for each quasar field. \citet{Bosman2020} serendipitously found a Ly$\alpha$ forest transmission in the narrow-band (NB) filter along a background galaxy of a $z\simeq5.8$ quasar. This suggests that the NB photometric search of the enhanced Ly$\alpha$ forest transmission around a quasar may be possible. Similarly, \citet{Mawatari2017} have utilised the NB filter to search for the large-scale excess Ly$\alpha$ forest absoprtion in the $z\simeq3.1$ protocluster region along the known background galaxies. If the NB photometric imaging is a viable alternative to spectroscopic Ly$\alpha$ forest tomography, it provides an economical strategy to survey a wider and more target fields, capitalizing on higher throughput and wider field of view of imaging than spectroscopy, which allows us to potentially examine the quasar active growth history of SMBHs using a statistically representative sample.

In this paper, we examine the observational requirements and feasibility of this ``photometric IGM tomography'' to map the quasar's ionizing light echoes in order to constrain the radiative growth history of a SMBH. We first introduce the concept and overall observing strategy in Section \ref{sec:concept}. Section \ref{sec:background_sources} estimates the expected number of background sources. Section \ref{sec:mock} examines the the scope of the photometric IGM tomographic technique using realistic mock observation and the reconstruction of 2D Ly$\alpha$ forest map based on a cosmological hydrodynamic simulation. Section \ref{sec:inference} introduces a fully Bayesian inference framework and demonstrates the constraining power for the quasar lifetime. Section \ref{sec:discussion} discusses caveats, possible extension using Subaru/PFS, VLT/MOONS, and Keck/DEIMOS and other possible applications of photometric IGM tomography. The conclusions are presented in Section \ref{sec:conclusion}.

We adopt a $\Lambda$CDM cosmology with $H_0=67.7\rm\,km\,s^{-1}\,Mpc^{-1}$, $\Omega_\Lambda=0.0.693$, and $\Omega_\Lambda=0.307$ \citep{Planck2015}. We use the AB magnitude system \citep{Oke1983}. We denote proper Mpc as pMpc (1\,pMpc corresponds to the light crossing time of  3.26\,Myr) and comoving Mpc as cMpc.

\section{Concepts}\label{sec:concept}

\subsection{Quasar light echoes: the accretion history of a SMBH}

\begin{figure}
\centering
	\includegraphics[width=0.95\columnwidth]{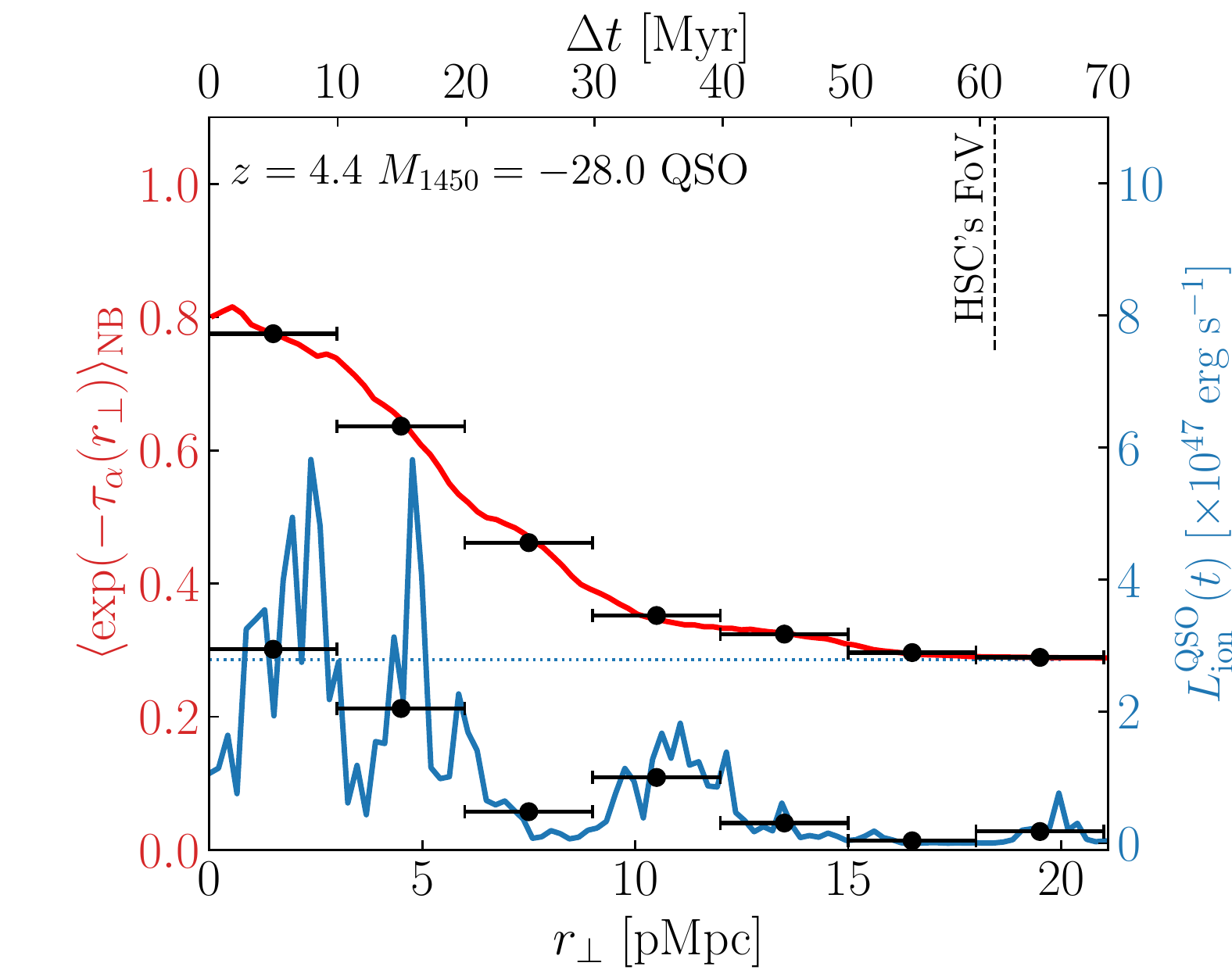}
    \caption{An example of the NB integrated Ly$\alpha$ forest transmission profile (red, left y-axis) as a function of impact parameter and the corresponding lightcurve (blue, right y-axis) of the quasar ionizing luminosity using a phenomelogical stochatic model in \citet{Bosman2020}. The impact parameter can be directly translated into time using $\Delta t=r_\perp/c$. We assume a luminous quasar at redshift of $z=4.4$ and UV magnitude of $M_{1450}=-28.0$. The figure demonstrates how the quasar lightcurve appears as a transverse proximity effect which can be mapped using IGM tomography. The errorbar represents the mean transverse separation between background galaxies (typically $\Delta r_\perp\sim3\rm\,pMpc$ as dicussed below), which determines the spatial resolution of the IGM tomography. This in turn sets the temporal resolution ($\Delta r_\perp/c\simeq 9.8\,\rm Myr$) for the quasar lightcurve constraint. This demonstrates that the light-echo tomography enables us to measure the lightcurve of a high-redshift quasar up to $\sim60\rm\,Myr$ baseline. }
    \label{fig:analytic}
\end{figure}

We first illustrate how IGM tomography can be used to recover the quasar light curve. The IGM at physical transverse and line-of-sight distance $r_\perp$ and $r_\parallel$ away from a quasar will be photoionized by the intense ionizing radiation emitted from the accretion disk of the central SMBH with the photoionization rate,
\begin{equation}
    \Gamma^{\rm QSO}_{\rm HI}(r_\parallel,r_\perp)=\frac{\alpha_Q\sigma_{912}}{3+\alpha_Q}\frac{\dot{N}_{\rm ion}^{\rm QSO}\left[-\Delta t(r_\parallel,r_\perp)\right]}{4\pi(r^2_\parallel+r^2_\perp)},\label{eq:quasar_photoionization}
\end{equation}
where $\dot{N}_{\rm ion}^{\rm QSO}(t)=L_{\rm ion}^{\rm QSO}(t)/(h\nu_L)$ ($L_{\rm ion}^{\rm QSO}$ is the quasar luminosity at 912\,\AA) is the ionizing photon production rate of the quasar emitted at time $t$, $\alpha_Q$ is the power-law index $\propto \nu^{-\alpha_Q}$ at $h\nu>13.6\rm\,eV$, and $\sigma_{912}$ is the photoionization cross section evaluated at the Lyman edge ($h\nu_L=13.6\,\rm eV$).
Here we define $t=0$ to be the cosmic time corresponding to when the radiation emitted by the quasar arrives on Earth (i.e. time at the quasar's redshift) and $t=-\Delta t$ means $\Delta t$ time in past relative to the time at the quasar's redshift. In Figure \ref{fig:analytic} we show an example lightcurve of the quasar ionizing luminosity using a phenomenological stochastic model employed in \citet{Bosman2020}.

\begin{figure*}
    \hspace*{-0.3cm}
    \includegraphics[width=0.85\columnwidth]{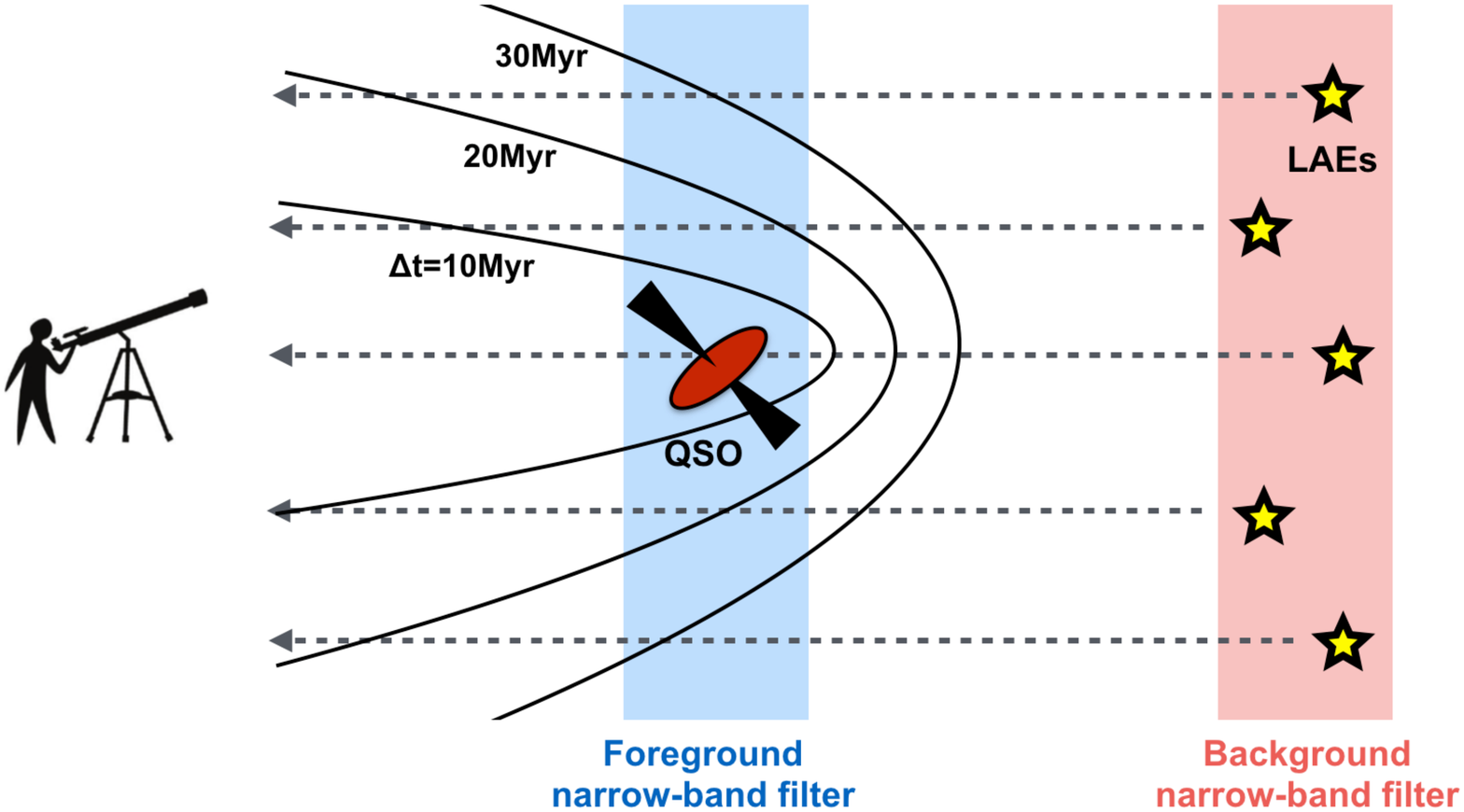}
    \hspace*{-0.23cm}
    \raisebox{0.05\height}{\includegraphics[width=1.25\columnwidth]{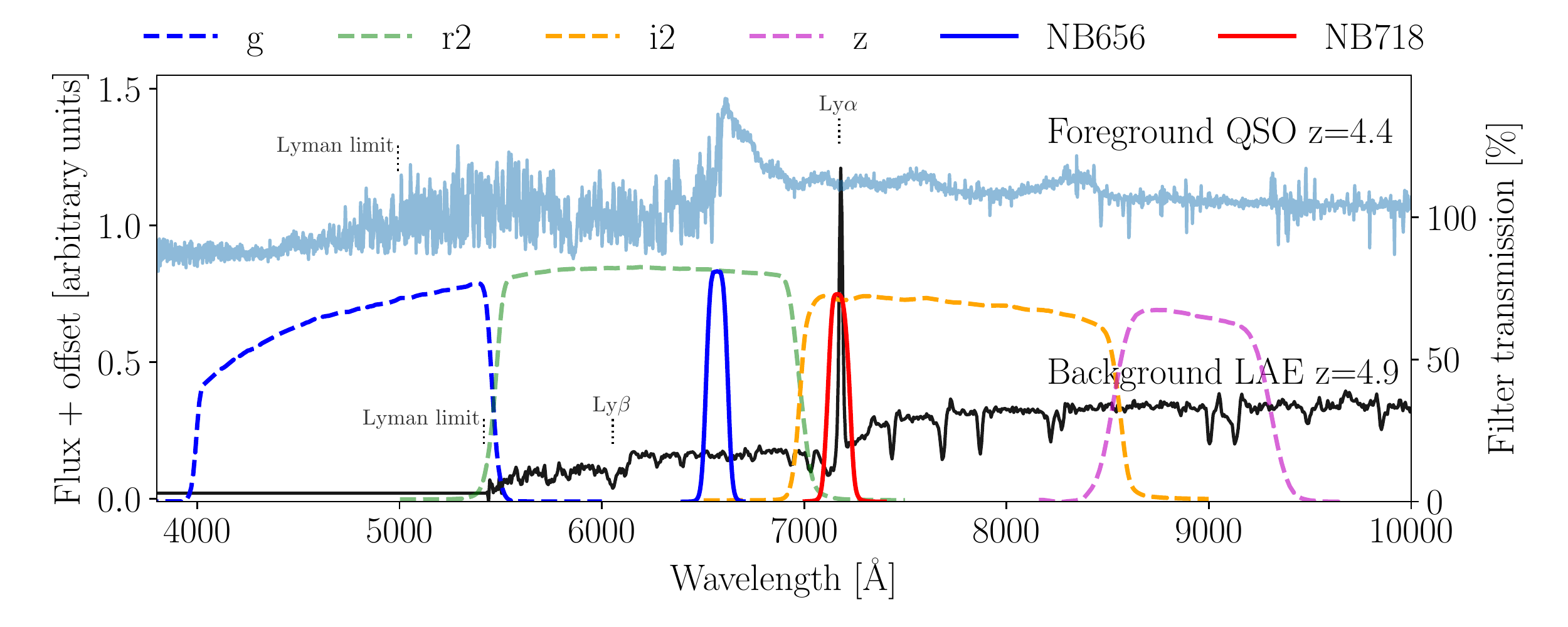}}
    \vspace{-0.3cm}
    \caption{({\bf Left}): Schematic illustration of the observing strategy for NB IGM tomography. We use a pair of NB filters. The red NB filter is used to identify the background galaxies using the NB excess due to the Ly$\alpha$ emission line (i.e. LAEs) and the blue NB filter is used to measure the transmitted Ly$\alpha$ forest fluxes around the foreground quasar towards the background galaxies. The time delay surfaces at $\Delta t$ are shown by the paraboloids. ({\bf Right}): Illustration of how a pair of NB filters of Subaru/HSC covers the foreground Ly$\alpha$ forest transmission at $z\simeq4.4$ (via NB656) and the Ly$\alpha$ emission line (via NB718) of a background LAE at $z\simeq4.9$ (black solid, based on \citet{Shapley2003} galaxy spectrum). An example foreground quasar spectrum at $z\simeq4.4$ is indicated by the blue solid line (based on a SDSS DR16 quasar spectrum from \citealt{Lyke2020}). The throughputs of BB filters are indicated by the coloured dashed lines. The UV continuum of a background LAE is measured by the $z$-band filter.}
    \label{fig:illustration}
\end{figure*}

This effect of quasar photoionization of the IGM can be traced by the Ly$\alpha$ forest absorption along background galaxies. Because the time of a photon to travel from quasar to a point in the IGM at $r_\parallel$ and $r_\perp$ is finite, each position of the IGM is influeneced by the quasar activity at different time in past with a time lag,
\begin{equation}
    \Delta t=\frac{(r^2_\parallel+r^2_\perp)^{1/2}-r_\parallel}{c}
\end{equation}
This defines the paraboloid surface of constant time lag \citep{Adelberger2004} as illustrated in Figure \ref{fig:illustration}. For a NB redshift slice centred at the quasar redshift, we can approximately set $r_{\parallel}\approx0$. The transverse distance from the quasar thus directly translates into the time of past quasar activity at $\Delta t\approx r_\perp/c$. As the bandwidth of the NB filter integrates a large segment of the IGM averaging over gas density fluctuations, we expect that the NB-integrated Ly$\alpha$ forest transmission $\langle \exp(-\tau_\alpha(r_\perp))\rangle_{\rm NB}$ follows as
\begin{align}
\langle\exp(-\tau_\alpha(r_\perp)\rangle_{\rm NB}
\approx\!\int^{\sqrt{r^2_\perp+L^2_{\rm NB}}}_{r_\perp}\langle\exp(-\tau_\alpha(r_\parallel,r_\perp)\rangle\frac{rdr}{L_{\rm NB}\sqrt{r^2-r_\perp^2}},
\end{align}
where $r=\sqrt{r^2_\parallel+r^2_\perp}$ is the radial distance from the quasar to a point in the IGM, $L_{\rm NB}=\frac{c\Delta\lambda_{\rm NB}}{2H(z_\alpha)\lambda_{\rm NB}}$ is a half of the IGM length averaged over the NB filter width with the full width at half maximum $\lambda\Delta\lambda_{\rm NB}$ and the central wavelength $\lambda_{\rm NB}$ corresponding to Ly$\alpha$ redshift $z_\alpha$, and $\langle\exp(-\tau_\alpha(r_\parallel,r_\perp)\rangle$ is the 3D mean Ly$\alpha$ forest transmission around a quasar \citep{Kakiichi2018,Bosman2020},
\begin{align}
&\langle \exp(-\tau_\alpha(r_\parallel,r_\perp)\rangle\approx \nonumber \\
&~~~~~~~~~\int d\Delta_b P_{\rm V}(\Delta_b)\exp\left[-\tau_0\Delta_b^\beta\left(1+\frac{\Gamma_{\rm HI}^{\rm QSO}(r_\parallel,r_\perp)}{\bar{\Gamma}_{\rm HI}}\right)^{-1}\right],
\end{align}
where $\beta=2-0.72(\gamma-1)$ with $\gamma$ being the slope of temperature-density relation $T=T_0\Delta_b^{\gamma-1}$, $P_{\rm V}(\Delta_b)$ is the density probability distribution function of the IGM overdensities $\Delta_b$, $\tau_0\simeq2.2(1+\chi_{\rm He})(\bar{\Gamma}_{\rm HI}/10^{12}{\rm\,s^{-1}})^{-1}(T_0/10^4{\rm\,K})^{-0.72}[(1+z)/5]^{9/2}$ is the Gunn-Peterson optical depth of Ly$\alpha$ forest at mean density and mean photoionization rate $\bar{\Gamma}_{\rm HI}$, and $\chi_{\rm He}$ is the fraction of electrons released by singly ionized helium ($\chi_{\rm He}\simeq0.0789$).

Figure \ref{fig:analytic} shows the NB-integrated Ly$\alpha$ forest transmission around a quasar as a function of impact parameter $r_\perp$ and the corresponding quasar lightcurve for a $z=4.4$ quasar with the UV magnitude $M_{\rm 1450}=-28.0$, assuming the NB filter width of $\Delta\lambda_{\rm NB}=100$\,\AA~($L_{\rm NB}\simeq4.8\,\rm pMpc$). The figure demonstrates the correlation between the lightcurve and the NB-integrated Ly$\alpha$ forest transmission around a quasar. The observed Ly$\alpha$ forest transmission profile is the coarse-grained version of the underlying quasar lightcurve smoothed over the scale of the NB filter width and the spatial sampling of the background galaxies. The horizontal errorbars represent an example mean separation between background galaxies, which determines the spatial resolution of the IGM tomography and the temporal resolution of the reconstructed quasar lightcurve. The direct correspondence between impact parameter $r_\perp$ and the time delay $\Delta t=r_\perp/c$ allows us to translate the measurement of the Ly$\alpha$ forest transmission at various impact parameters into the lightcurve constraint. The rise of quasar activity from $\Delta t\simeq-20\rm\,Myr$ is seen in the Ly$\alpha$ forest transmission profile at $r_\perp\lesssim6\rm\,pMpc$. As well, the quasar burst at $\Delta t\simeq-35\rm\,Myr$ is seen as an extended tail in the Ly$\alpha$ forest transmission profile as an excess transmission at $r_\perp\simeq11\rm\,pMpc$.

This makes it possible to measure the impact of a variable quasar lightcurve on the IGM using the Ly$\alpha$ forest tomography along background galaxies. As the time sampling reflects the travel time between two points in space, {\it by spatially mapping this `light-echo', one can translate the Mpc-scale spatial information of the IGM into a Myr-timescale time-domain constraint on the quasar lightcurve of an individual SMBH over the baseline of $\sim60\rm\,Myr$}. This provides an observational tool to measure the growth history of a SMBH over the timescale (one e-folding $t_{\rm sal}=45\rm\,Myr$) required to assemble a substantial fraction of its mass.

\subsection{Double narrow-band IGM tomography}

To implement the IGM tomography of quasar light-echoes using photometry, we can use a pair of NB filters to map the Ly$\alpha$ forest transmission around an individual quasar, which we refer to as the `double NB technique'. The experimental configuration is illustrated in Figure \ref{fig:illustration}. In this technique, we select a pair of NB filters: ({\it i}) a blue filter corresponds to the redshift of the foreground quasar and ({\it ii}) a red filter corresponds to the redshift of the background sources. We first identify the background Ly$\alpha$ emitters (LAEs) using the standard NB technique \citep[e.g.][]{Ono2021} to be used as background sources for IGM tomography. Along these background LAEs, deep exposures in the foreground blue NB filter measures the transmitted Ly$\alpha$ forest flux within the NB filter width. Broad-band (BB) imaging will be used to measure the UV continuum level of the background LAEs. The flux ratio (or magnitude difference) between the inferred Ly$\alpha$ forest flux and the observed flux within the foreground NB filter provides a measure of the Ly$\alpha$ forest transmission at the redshift slice of the foreground quasar.

This double NB technique for IGM tomography provides a couple of advantages over the conventional full spectroscopic tomographic method to quantify the transmission through the sightlines of the background galaxies. First, the double NB method circumvents the need for an expensive spectroscopic follow-up campaign: both to spectroscopically confirm the background galaxy candidates selected by pre-imaging and to obtain the deep spectroscopic data to detect the UV continua and faint Ly$\alpha$ forest transmissions along the background galaxies as required for Ly$\alpha$ forest tomography in the traditional method. Second, imaging can typically go deeper relative to spectroscopy due to the higher throughput. HSC NB filters have the end-to-end throughput of $\sim60\,\%$ whereas even the most senstive multislit spectrographs have throughput below $25\,\%$. The photometric method therefore can be more sensitive to faint Ly$\alpha$ forest transmission along the background galaxies. Although the reconstructed Ly$\alpha$ forest transmission map is 2D for the NB tomographic method and the line-of-sight information is averaged over the width of NB filter, because of the larger field-of-view of wide-field imagers compared to those typical of wide-field multi-object spectrographs, NB IGM tomography can outperform the spectroscopic method by surveying much larger area of sky in a single pointing.

\subsection{Filter set}

\begin{table*}
	\centering
	\caption{Possible HSC NB filter combinations for double NB photometric Ly$\alpha$ forest tomographic technique. The checkmark indicates a suitable filter combination so that the Ly$\alpha$ redshift of foreground NB filter (rows) matches with the Ly$\alpha$ forest region of the background LAEs located by the background NB filter (columns).}
	\label{table:filterset}
\begin{tabular}{@{\hskip 1mm}c@{\hskip 1mm}@{\hskip 1mm}c@{\hskip 1mm}|c@{\hskip 1mm}c@{\hskip 1mm}c@{\hskip 1mm}c@{\hskip 1mm}c@{\hskip 1mm}c@{\hskip 1mm}c@{\hskip 1mm}c@{\hskip 1mm}c@{\hskip 1mm}c@{\hskip 1mm}c@{\hskip 1mm}c@{\hskip 1mm}c@{\hskip 1mm}c@{\hskip 1mm}c@{\hskip 1mm}c}
\hline\hline
                & bg. filter & NB391 & NB395 & NB400 & NB430 & NB468 & NB497 & NB506 & NB515 & NB527 & NB656 & NB718 & NB816 & NB921 & NB926 & NB973 & NB1010 \\
fg. filter & $z_{\rm Ly\alpha}$  & 2.22  & 2.25  & 2.29  & 2.54  & 2.85  & 3.09  & 3.16  & 3.24  & 3.33  & 4.39  & 4.90  & 5.71  & 6.57  & 6.62  & 7.00  & 7.31   \\
\hline
\rowcolor{Gray}
NB387  & 2.18  & $\checkmark$ & $\checkmark$ & $\checkmark$ & $\checkmark$ &    &    &    &    &    &    &    &    &    &    &    &     \\
NB391   & 2.22  &    & $\checkmark$ & $\checkmark$ & $\checkmark$ &    &    &    &    &    &    &    &    &    &    &    &     \\
\rowcolor{Gray}
NB395  & 2.25  &    &    & $\checkmark$ & $\checkmark$ & $\checkmark$ &    &    &    &    &    &    &    &    &    &    &     \\
NB400  & 2.29  &    &    &    & $\checkmark$ & $\checkmark$ &    &    &    &    &    &    &    &    &    &    &     \\
\rowcolor{Gray}
NB430  & 2.54  &    &    &    &    & $\checkmark$ & $\checkmark$ & $\checkmark$ &    &    &    &    &    &    &    &    &     \\
NB468  & 2.85  &    &    &    &    &    & $\checkmark$ & $\checkmark$ & $\checkmark$ & $\checkmark$ &    &    &    &    &    &    &     \\
\rowcolor{Gray}
NB497  & 3.09  &    &    &    &    &    &    & $\checkmark$ & $\checkmark$ & $\checkmark$ &    &    &    &    &    &    &     \\
NB506  & 3.16  &    &    &    &    &    &    &    & $\checkmark$ & $\checkmark$ &    &    &    &    &    &    &     \\
\rowcolor{Gray}
NB515  & 3.24  &    &    &    &    &    &    &    &    & $\checkmark$ &    &    &    &    &    &    &     \\
NB527  & 3.33  &    &    &    &    &    &    &    &    &    &    &    &    &    &    &    &     \\
\rowcolor{Gray}
NB656  & 4.39  &    &    &    &    &    &    &    &    &    &    & $\checkmark$ &    &    &    &    &     \\
NB718  & 4.90   &   &    &    &    &    &    &    &    &    &    &    & $\checkmark$ &    &    &    &     \\
\rowcolor{Gray}
NB816  & 5.71  &    &    &    &    &    &    &    &    &    &    &    &    & $\checkmark$    & $\checkmark$ &    &     \\
NB921  & 6.57  &    &    &    &    &    &    &    &    &    &    &    &    &    & $\checkmark$ & $\checkmark$ & $\checkmark$  \\
\rowcolor{Gray}
NB926  & 6.62  &    &    &    &    &    &    &    &    &    &    &    &    &    &    & $\checkmark$ & $\checkmark$  \\
NB973  & 7.00  &    &    &    &    &    &    &    &    &    &    &    &    &    &    &    & $\checkmark$  \\
\hline\hline
\end{tabular}
\end{table*}

\begin{figure}
	\vspace{-0.5cm}
	\includegraphics[width=\columnwidth]{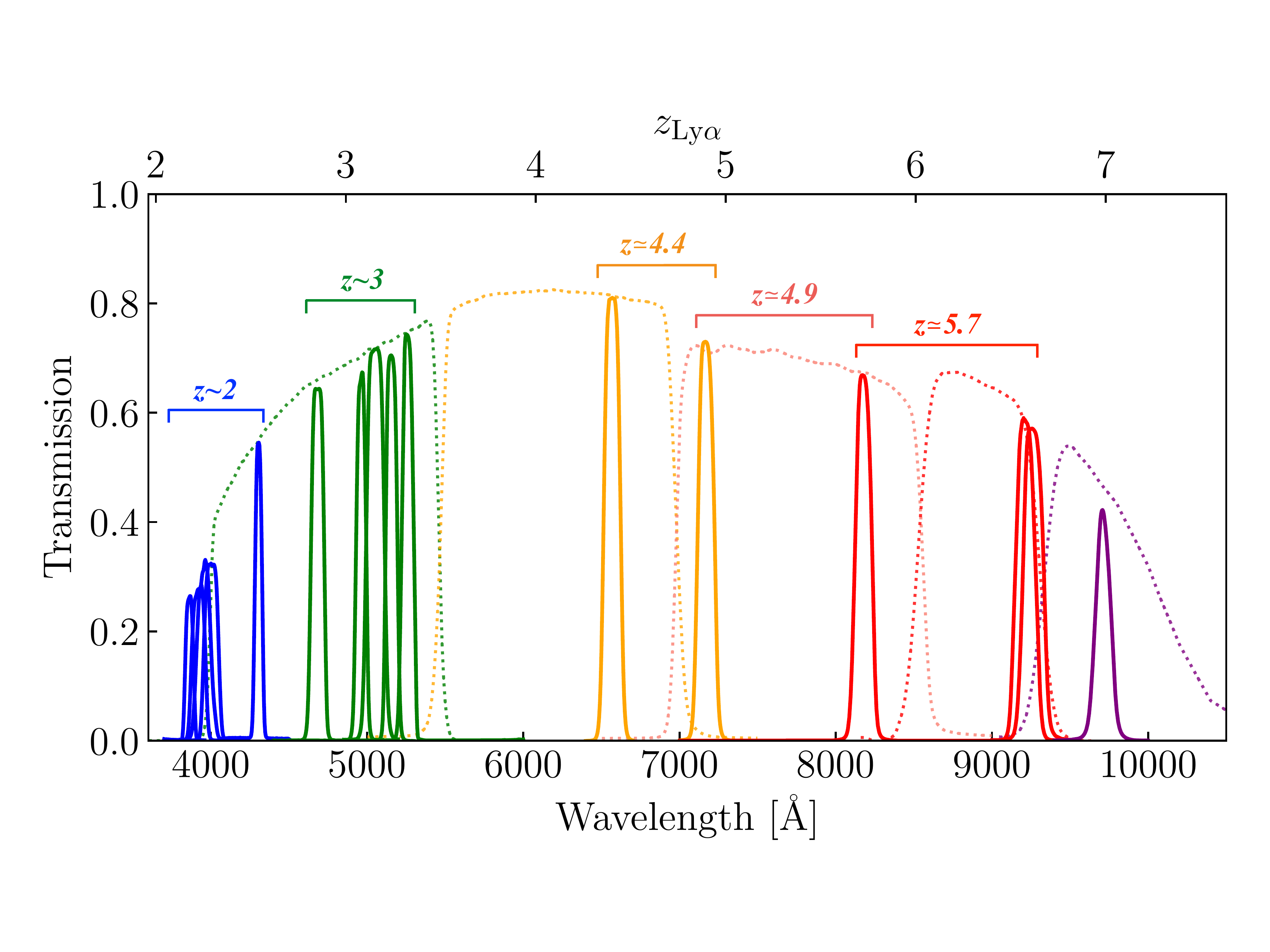}
	\vspace{-1.0cm}
    \caption{The possible combinations or pairs of HSC NB filters for photometric Ly$\alpha$ forest tomography. The filter transmissions of NB filters (solid) including CCD quantum efficiency and transmittance through the dewar window and the primary focus unit of the HSC (from left to right: (blues) NB387, NB391, NB395, NB400, NB430, (greens) NB468, NB497, NB506, NB515, NB527, (yellows) NB656, NB718, (reds) NB816, NB921, NB926, (purple) NB973). The filter transmissions for the BB $g,r2,i2,z,Y$ filters are also indicted by the dotted curves.}\label{fig:filterset}
\end{figure}

To study quasar light echoes with the double NB IGM tomography, the foreground NB filter needs to cover the rest-frame wavelength of Ly$\alpha$ line of the quasar at redshift $z_Q$. For convenience, we assume the central wavelength of the foreground NB filter coincides with the quasar's Ly$\alpha$ redshift $\lambda_{\rm NB}=\lambda_\alpha(1 + z_Q)$. Then, to measure the Ly$\alpha$ forest transmission around the quasar, the Ly$\alpha$ forest range between Ly$\alpha$ and Ly$\beta$ lines of the background galaxies should be covered by the the foreground NB filter, requiring the redshift $z_{\rm bkg}$ of a background galaxy to be
\begin{equation}
 z_Q<z_{\rm bkg}<\lambda_\alpha(1+z_Q)/\lambda_\beta-1,
\end{equation}
where $\lambda_\alpha=1216$~\AA~and $\lambda_\beta=1026$~\AA~are the Ly$\alpha$ and Ly$\beta$ wavelengths. For the double NB technique, we locate the background galaxies with a (background) NB filter redward of the foreground NB filter. Table \ref{table:filterset} shows all the possible pairs of NB filters for Subaru/HSC. The dense wavelength separations of the NB filters mean that we can apply this double NB IGM tomography to all NB filters expect for NB527, covering Ly$\alpha$ redshift from $z\simeq2.18$ to $7.0$. Note that when designing a realistic survey, one should also take into account the filter widths of the foreground and background NB filter and may consider using shorter Ly$\alpha$ forest range to avoid possible compliations in the intrinsic galaxy spectral energy distribution (SED) near Ly$\alpha$ and Ly$\beta$ lines. Table \ref{table:filterset} should be regarded as the inclusive list of possible NB filter pairs.

We highlight the interesting filter pair combinations in Figure \ref{fig:filterset}. Particularly interesting combinations are:
\smallskip

\noindent For $z\sim2$,
\[
\indent\mbox{foreground NB filters}=
\begin{cases}
\rm NB387 \\
\rm NB391 \\
\rm NB395 \\
\rm NB400 \\
\end{cases}
(z\simeq2.18-2.29)
\]
\indent AND
\[
\indent \mbox{background NB filter}={\rm NB430}~~~(z\simeq2.54).
\]
\smallskip

\noindent For $z\sim3$,
\[
\indent \mbox{foreground NB filter set}=
\begin{cases}
\rm NB497 \\
\rm NB506 \\
\rm NB515 \\
\end{cases}
(z\simeq3.09-3.24)
\]
\indent AND
\[
\indent \mbox{background NB filter} ={\rm NB527}~~~(z\simeq3.33).
\]
\smallskip

\noindent For $z\sim4$,
\[
\indent \mbox{foreground NB filter}={\rm NB656}~~~(z\simeq4.4)
\]
\indent AND
\[
\indent \mbox{background NB filter}={\rm NB718}~~~(z\simeq4.9).
\]
\smallskip

\noindent For $z\sim5$,
\[
\indent \mbox{foreground NB filter}={\rm NB718}~~~(z\simeq4.9)
\]
\indent AND
\[
\indent \mbox{background NB filter}={\rm NB816}~~~(z\simeq5.7).
\]
\smallskip

\noindent For $z\sim6$,
\[
\indent \mbox{foreground NB filter}={\rm NB816}~~~(z\simeq5.7)
\]
\indent AND
\[
\indent \mbox{background NB filters}=
\begin{cases}
\rm NB921 \\
\rm NB926 \\
\end{cases}
(z\simeq6.6).
\]
\smallskip

For $z\sim2$ and $z\sim3$ filter combinations, the densely populated HSC NB filters make it possible to perform a pseudo-3D photometric IGM tomography coarsely sampled along the line of sight direction averaged over $\sim35h^{-1}\rm cMpc$, corresponding to a typical $\sim100$\,\AA~NB width across the entire 1.5 deg field of view in diameter ($\sim100h^{-1}\rm cMpc$).
At $z\sim4-6$, while the mapping is limited to 2D, the imaging's higher sensitivity to fainter Ly$\alpha$ forest transmission allows us to examine the IGM tomography at higher redshifts than that achievable spectroscopically with 8-10m class telescopes. As the surface number density of background galaxies
defines the spatial resolution of the IGM tomographic map, the availability of two background NB filters, NB921 and NB926, for $z\simeq5.7$ NB816 tomography allows us to increase the density of background LAEs. This effective increase in the survey volume of background LAEs may be more efficient than going deeper with a single filter, especially at the redshift where the number density of observable LAEs is rapidly diminishing due to the effect of reionization. At $z>6.6$, while the filter combination (e.g. NB921 for foreground NB filter and NB973 for background NB filter) permits attempting even higher-redshift IGM tomography, the scarcity of LAEs at $z>7$ would make it impractical for LAEs to be used as background sources. One might require alternative selection of background sources such as using the $\rm H\beta+[\OIII]$ systems selected by JWST NIRCam grism spectroscopy to make IGM tomography in the reionization era possible.

\begin{figure*}
	\includegraphics[width=0.33\textwidth]{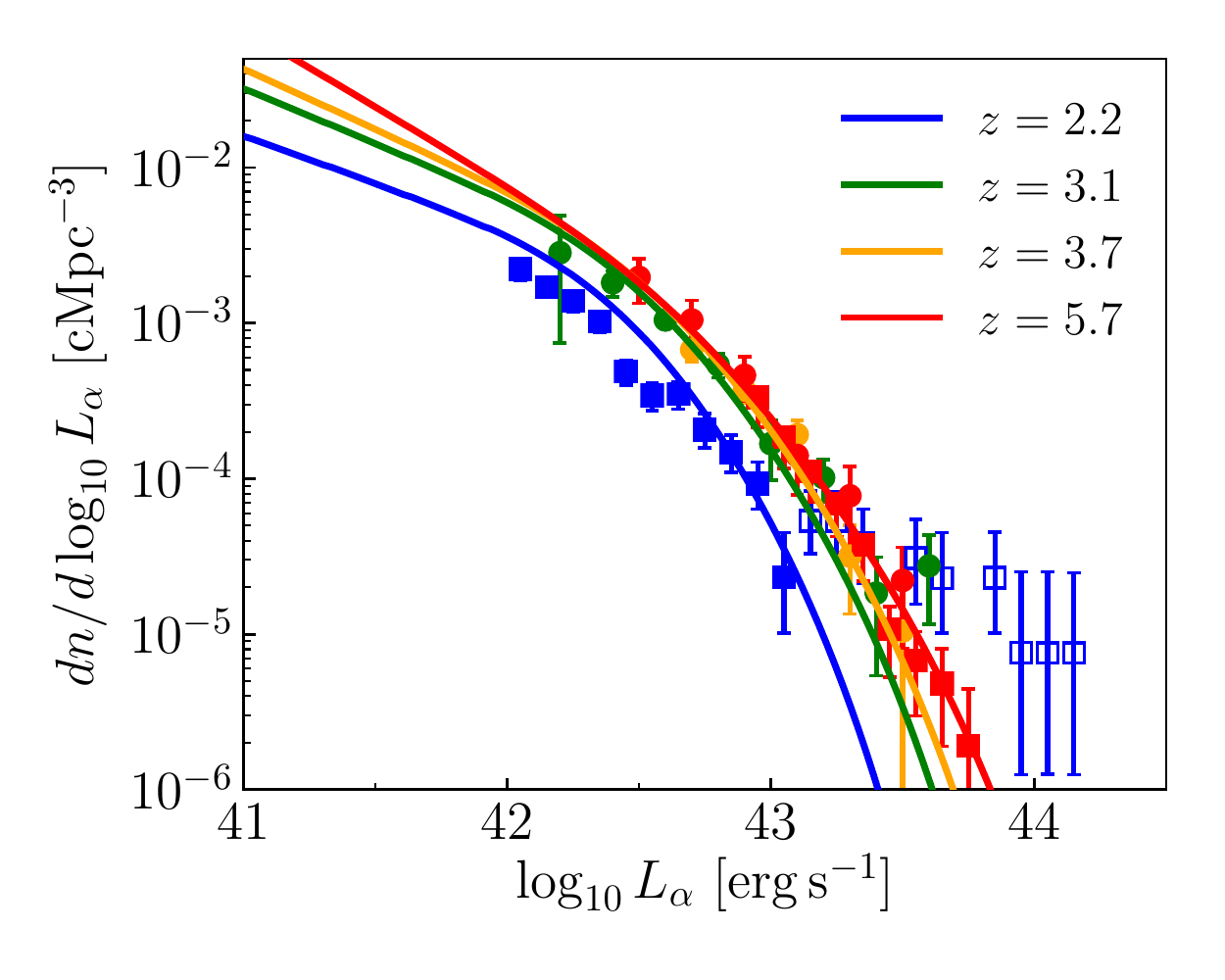}
	\includegraphics[width=0.33\textwidth]{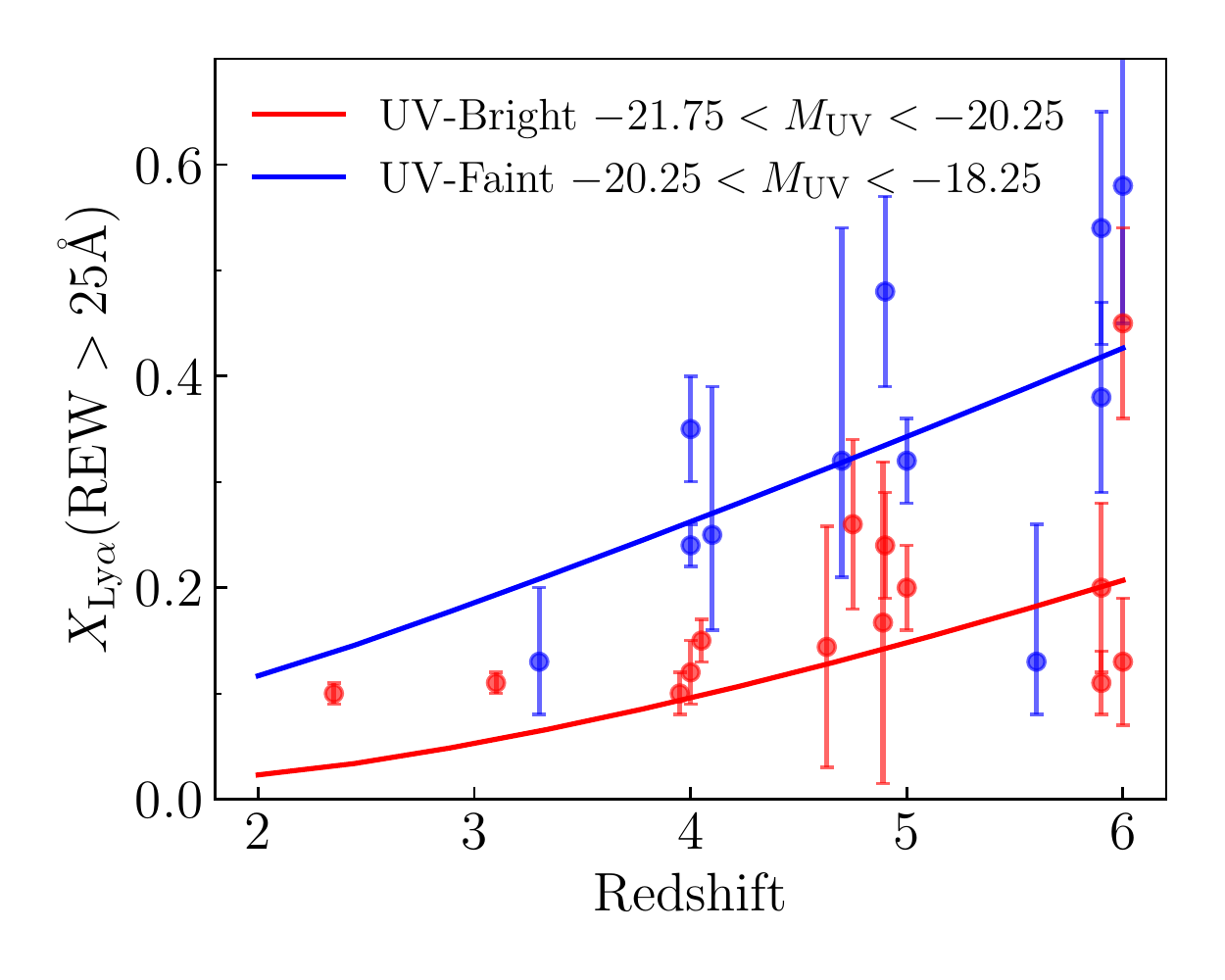}
	\includegraphics[width=0.33\textwidth]{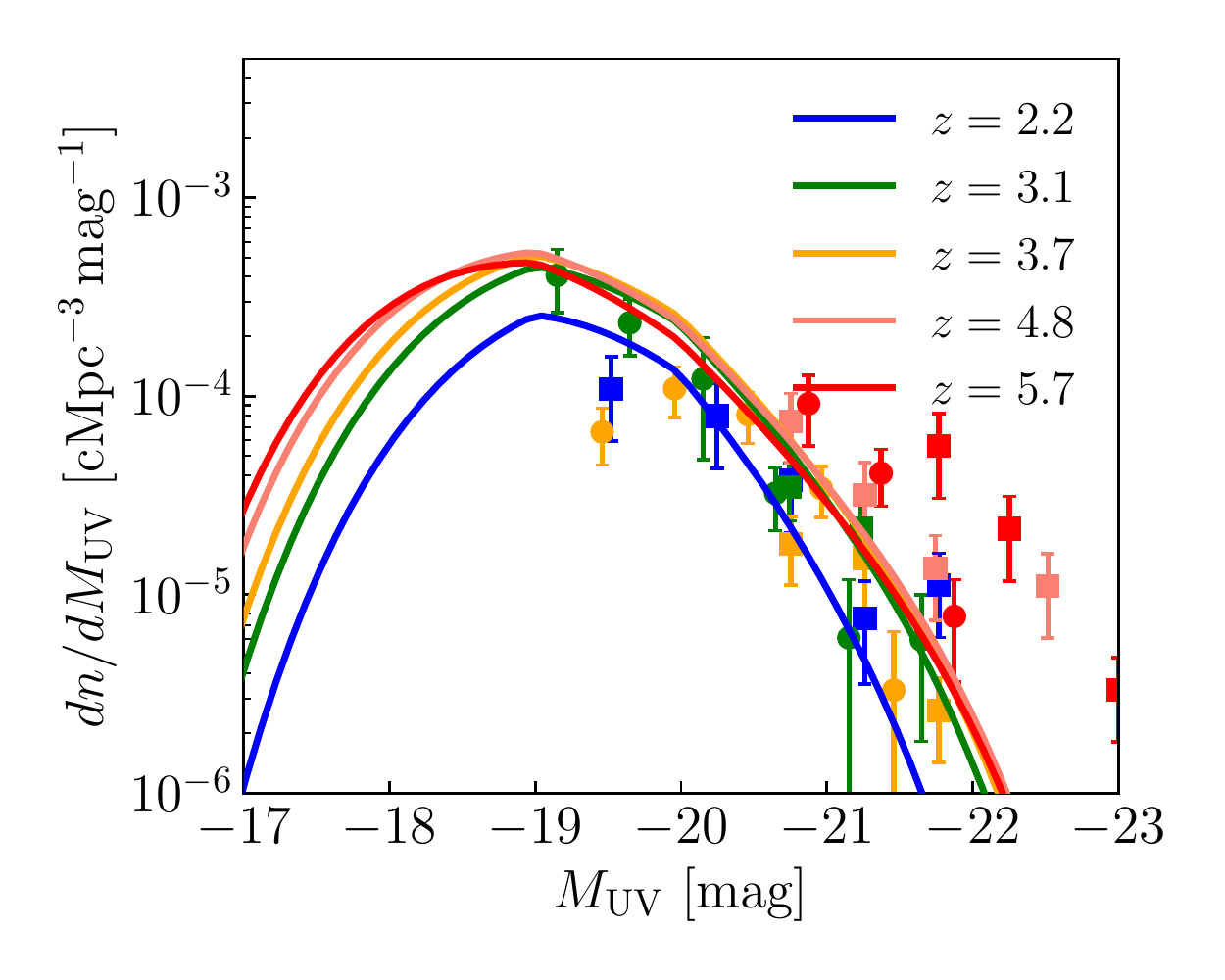}
	\vspace{-0.5cm}
    \caption{{\bf (Left):} Comparison between the model Ly$\alpha$ luminosity functions of LAEs (coloured lines) and observations at $z=2.2$ (blue), 3.1 (green), 3.7 (yellow), and 5.7 (red) \citep{Ouchi2008,Konno2016,Konno2018}. The open symbols indicate the luminosity bins likely contaminated by AGN reported by \citet{Konno2018}. {\bf (Middle):} Comparison between the model Ly$\alpha$ fractions of UV-continuum selected galaxies (model: UV-bright $-21.75<\Muv<-20.25$ (red) and UV-faint $-21.75<\Muv<-20.25$ (blue)) and observations  \citep[red and blue circles for UV-bright and -faint samples: ][]{Stark2010,Stark2011,Mallery2012,Curtis-Lake2012,Cassata2015,deBarros2017,Arrabal-Haro2018,Kusakabe2020}. {\bf (Right):} Comparison between the model UV luminosity functions of LAEs (solid) and observations at $z=2.2$ (blue), 3.1 (green), 3.7 (yellow), 4.8 (salmon), and 5.7 (red) (circles, \citealt{Ouchi2008}; squares, \citealt{Santos2021}).}
    \label{fig:LAE}
\end{figure*}

\section{Background sources}\label{sec:background_sources}

\subsection{Requirement for background sources}

The spatial resolution of IGM tomography is determined by the number density of background sources.  This in turn determines the required survey depth for the IGM tomography at a desired spatial resolution. In double NB IGM tomography, the background sources need to be LAEs with bright UV continua so that the ratio between the transmitted Ly$\alpha$ forest fluxs and infered continua, i.e. Ly$\alpha$ forest transmission, can be measured from the foreground NB filter and the BB filters. This is different from the requirement for the background sources for conventional spectroscopic IGM tomography \citep{Lee2014a,Lee2014b,Lee2018,Newman2020}, for which one can use all star-forming galaxies selected by Lyman-break technique with measurable UV contina regardless of their Ly$\alpha$ lines. Double NB tomography can therefore only use a subset of star-forming galaxies compared to those used for the spectroscopic IGM tomography. However, because double NB tomography does not require detecting the UV continuum spectroscopically, much fainter objects can be used as background sources. In order to compare pros and cons of the two tomographic methods, we need to treat both LAEs and Lyman-break galaxies (LBGs) in the same framework.

\subsection{Model and observations}

Observations indicate that LAEs are a subset of star-forming galaxies with young ages, low stellar mass, and little dust \citep[e.g.][for recent review]{Ouchi2020}. Following the \citet{Dijkstra2012}, we construct an empirical model matched to the LAE and LBG luminosity functions across $z\sim2-6$. The central model quantity is the probability of a galaxy showing a Ly$\alpha$ emission line with a  rest-frame equivalent width (REW) at a given UV magnitude $\Muv$ (hereafter the REW-PDF), which can be modelled as
\begin{equation}
P({\rm REW}|M_{\rm UV})=\mathcal{F}\exp\left(-\frac{\rm REW}{{\rm REW_c(M_{\rm UV})}}\right)  ,
\end{equation}
where ${\rm REW_c}=23+7(M_{\rm UV}+21.9)+6(z-4)$ according to the best-fit model of \citet{Dijkstra2012}. The pre-factor\footnote{\citet{Dijkstra2012} choose a constant numerical factor $0.44$ instead of $0.5(z/5.7)$. This revision was necessary to better match with the updated measurement of Ly$\alpha$ fraction of LBGs.} $\mathcal{F}$ is given by $\mathcal{F}=\frac{0.5(z/5.7)}{{\rm REW_c(\Muv)}}\left[\exp\left(\frac{{\rm REW}_{\rm min}}{\rm REW_c(\Muv)}\right)-\exp\left(\frac{{\rm REW}_{\rm max}}{\rm REW_c(\Muv)}\right)\right]^{-1}$ where  ${\rm REW}_{\rm min}=-20+6(\Muv+21.5)^2$\,\AA~for $-21.5\le\Muv\le-19.0$, ${\rm REW}_{\rm min}=20.0$\,\AA~for $\Muv<-21.5$, and ${\rm REW}_{\rm min}=17.5$\,\AA~for $\Muv>-19.0$. Since not all galaxies show Ly$\alpha$ emission, we define the REW-PDF to be normalized to the total fraction of galaxies with a UV magnitude $\Muv$ showing Ly$\alpha$ line in emission, i.e. $\int P({\rm REW}|M_{\rm UV})d{\rm REW}=0.5(z/5.7)$ where the numerical factor is a model parameter empirically chosen to match the observations below.

In this model, the Ly$\alpha$ luminosity function of LAEs can be expressed in terms of the REW-PDF and the UV luminosity function of star-forming galaxies $dn/d\Muv$ (for which we use the \citet{Bouwens2021} best-fit Schechter functions at $z\sim2-10$),
\begin{equation}
    \frac{dn_{\rm \scriptscriptstyle LAE}}{dL_\alpha}=\int^\infty_{-\infty} P_{\rm obs}(L_\alpha|\Muv)\frac{dn}{d\Muv}d\Muv,
\end{equation}
where $P_{\rm obs}(L_\alpha|\Muv)$ is the probability of an object with an UV magnitude $\Muv$ to be observed as a LAE with Ly$\alpha$ luminosity $L_\alpha$, which is given by $P_{\rm obs}(L_\alpha|\Muv)=\Theta({\rm REW}-{\rm REW}_{\rm cut})P({\rm REW}|\Muv)\left|\frac{d{\rm REW}}{dL_\alpha}\right|$ and $\left|\frac{d{\rm REW}}{dL_\alpha}\right|=\frac{\lambda_\alpha}{\nu_\alpha}\left(\frac{\lambda_{\rm 1600}}{\lambda_\alpha}\right)^{\beta+2}L_{\nu,{1600}}$ with $L_{\nu,{1600}}$ being the specific UV luminosity at $\lambda_{1600}=1600$\,\AA~and $\nu_\alpha$ and $\lambda_\alpha$ being the rest-frame frequency and wavelength of Ly$\alpha$ line.
We assume the UV continuum slope to be $\beta=-1.8$. The effect of NB selection (i.e. $\rm REW>REW_{\rm cut}=25$\,\AA) is included with the the heaviside step function $\Theta({\rm REW}-{\rm REW}_{\rm cut})$.

The Ly$\alpha$ fraction $X_\alpha(>{\rm REW}|M_{\rm UV}^{\rm min},M_{\rm UV}^{\rm max})$ of UV-continuum selected galaxies in a UV magnitudes bin $M_{\rm UV}^{\rm min}<\Muv<M_{\rm UV}^{\rm max}$ can similarly be expressed in terms of the REW-PDF and UV luminosity function,
\begin{align}
&X_\alpha(>{\rm REW}|M_{\rm UV}^{\rm min},M_{\rm UV}^{\rm max})= \nonumber \\
&~~~~~~~~~~~~
\frac{1}{n_{\rm UV}}\int_{M_{\rm UV}^{\rm min}}^{M_{\rm UV}^{\rm max}}d\Muv\frac{dn}{d\Muv}
\int_{\rm REW}^{\infty}d{\rm REW}\,P({\rm REW}|\Muv).
\end{align}
where $n_{\rm UV}=\int_{M_{\rm UV}^{\rm min}}^{M_{\rm UV}^{\rm max}}\frac{dn}{d\Muv}d\Muv$ is the number density of galaxies with the UV magnitude interval.

We can also express the UV luminosity function of LAEs with Ly$\alpha$ equivalent width ($\rm REW>\rm REW_{\rm cut}=25$\,\AA) as
\begin{equation}
    \frac{dn_{\rm\scriptscriptstyle LAE}}{d\Muv}=\int_{{\rm REW}_{\rm cut}}^\infty P({\rm REW}|\Muv)d{\rm REW}\times\frac{dn}{d\Muv}.\label{eq:UV_LF_LAE}
\end{equation}

In Figure \ref{fig:LAE}, we show the comparison of the emprical model with observations. It confirms that the empirical model agrees well with the measurements of Ly$\alpha$ luminosity functions of LAEs, Ly$\alpha$ fraction of UV-continuum selected galaxies, and the UV luminosity function of LAEs, justifying the use of the empirical model to estimate a realistic expected number density of background galaxies that satisfy the requirement of the UV and Ly$\alpha$ luminosities for the NB IGM tomography. At the bright-end of the Ly$\alpha$ luminosity function ($L_\alpha\gtrsim10^{43}\rm erg\,s^{-1}$) and UV luminosity function ($\Muv\lesssim-21$) of LAEs, the model appears to deviate from the observations. This is however likely due to AGN contamination to the luminosity functions at the bright-end \citep{Konno2016,Ono2018,Bowler2021}.
Thus, we do not consider this apparent mismatch to be an obvious shortcoming for estimating the background LAE density.
For IGM tomography, any type of background sources (AGN and galaxies) is sufficient. Since the observations indicates a higher density of bright LAEs, the empirical model should give a conservative lower limit for the background LAE density.

\subsection{Background source counts}

In order to use LAEs as background sources for IGM tomography, we require them to also be detected in the UV continuum filter. Thus, the relevant quantity is the surface number density of background LAEs with the UV magnitudes $<\Muv^{\rm lim}$, which is the integral over the UV luminosity function of LAEs,
\begin{equation}
    \Sigma_{\rm \scriptscriptstyle LAE}(<M^{\rm lim}_{\rm UV})=\int_{z_{\rm min}}^{z_{\rm max}}\left|\frac{dl_p}{dz}\right|(1+z)^3\int_{-\infty}^{\Muv^{\rm lim}}\frac{dn_{\rm\scriptscriptstyle LAE}}{d\Muv}d\Muv,\label{eq:sightline_density}
\end{equation}
where $|dl_p/dz|=c/[H(z)(1+z)]$, and $z_{\rm min}$ and $z_{\rm max}$ are defined from the FWHM of NB filter transmission curve of the background filter. The corresponding apparent UV magnitude is assumed to be $m_{\rm UV}=M_{\rm UV}+5\log_{10}(d_{\rm L}(z)/10{\rm\,pc})-2.5\log_{10}(1+z)$. Note that the sources of interest are selected via their NB excess (i.e. $\rm REW>25\rm\,\A$, $m_{\rm UV}-m_{\rm NB}>0.26$), which sets the required NB depth for a chosen $m_{\rm uv}^{\rm lim}$ at a given redshift.

\begin{figure}
	\includegraphics[width=1.05\columnwidth]{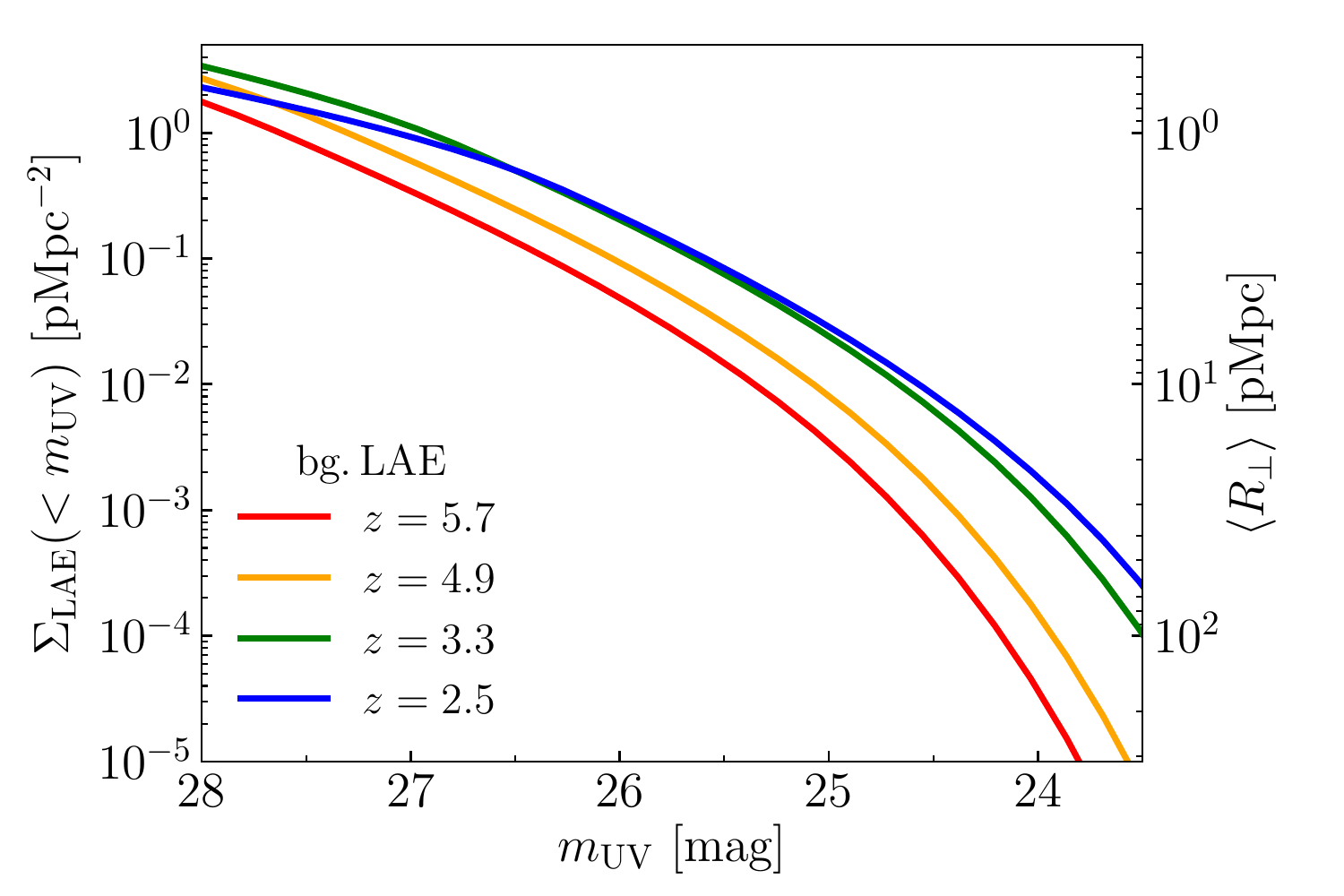}
	\vspace*{-0.5cm}
    \caption{The expected surface densities of LAEs identified via NB430 (red, $z=2.5$), NB527 (green, $z=3.3$), NB718 (yellow, $z=4.9$), NB816 (red, $z=5.7$) filters as a function of the limiting UV magnitudes. The right y-axis indicates the corresponding mean transverse separation of the background LAEs, $\langle R_\perp\rangle=\Sigma_{\rm LAE}^{-1/2}$.}
    \label{fig:surface_density}
\end{figure}

Figure~\ref{fig:surface_density} shows the surface number density of background LAEs as a function of their apparent UV magnitudes assuming NB selection with $\rm REW>25$\,\AA. As the background LAE surface number density ultimately sets the angular resolution of the IGM tomographic map, it is useful to compute a fitting formula; for $z=4.9$ background LAEs, we find
\begin{equation}
    \langle R_\perp\rangle\equiv\Sigma_{\rm LAE}^{-1/2}\approx 1.63\times10^{[(m_{\rm uv}/26.66)^{-9.52}-1]}\,{\rm pMpc}.
\end{equation}
This gives the typical spatial tomographic resolution of $\langle R_\perp\rangle\approx 1.9,\,3.0,\,5.5$\,pMpc for $m_{\rm uv}=26.5,\,26.0,\,25.5\rm\,mag$ respectively. The fitting formula for other background LAEs selected via different NB filters are shown in Table~\ref{tab:fit}, which are accurate to
$\sim10\%$ over the apparent UV magnitude range of $23.5<m_{\rm uv}<28.0$.

This directly translates to the time resolution for the quasar light-echoes,
\begin{equation}
    \Delta t\sim\frac{\langle R_\perp\rangle}{c}\approx 5.32\times10^{[(m_{\rm uv}/26.66)^{-9.52}-1]}\,{\rm Myr}.\label{eq:time_resol}
\end{equation}
For an approximately $\sim1-10\rm\,Myr$ time resolution, we require UV continuum depth of roughly $m_{\rm uv}\sim25.5-26.5$ mag to map out the quasar light echoes with NB IGM tomography using LAEs as background sources.
Thus, the observational requirement for the BB imaging
depth covering the UV-continuum of the background LAEs is set by the
desired spatial/temporal sampling which relates to the underlying transverse proximity effect/lightcurve structure
that experiment can resolve (see Figure \ref{fig:analytic}) as well as the overall extent of the detectable light echo signal.

\subsection{Figure-of-merit: narrow-band vs spectroscopic tomography}

\begin{table}
    \centering
    \caption{The best-fit values for the fitting formula $\langle R_{\perp}\rangle=A\times10^{[(m_{\rm uv}/m_0)^\gamma-1}]$ for various background LAEs. The results are accurate to within $\sim10\%$. The conversion between absolute and apparent UV magnitudes are assumed to be $m_{\rm UV}=M_{\rm UV}+5\log_{10}(d_{\rm L}(z)/10{\rm\,pc})-2.5\log_{10}(1+z)$.}
    \label{tab:fit}
    \begin{tabular}{llll}
    \hline\hline
         Filter & $A$ $[\rm pMpc]$  & $m_0$ $[\rm mag]$ & $\gamma$ \\
    \hline
         NB430  & 1.38 & 26.58        & $-7.90$    \\
         NB527  & 0.97 & 27.00        & $-7.82$    \\
         NB718  & 1.63 & 26.66        & $-9.52$    \\
         NB816  & 2.43 & 26.53        & $-10.45$   \\
     \hline\hline
    \end{tabular}
\end{table}

It is interesting to compare the pros and cons
of narrow-band and spectrosocpic tomographic techniques. We use the mean transverse resolution of background galaxies and the field-of-view as a figure-of-merit. Exact comparison of the two techniques per telescope time is difficult because the instruments (e.g. Subaru/HSC, Keck/DEIMOS or LRIS, Magellan/IMACS) suitable for the two techniques are typically installed on different telescopes.
In addition, practical considerations such as mask design and the number of available slits imply
that a real-world comparison would  need to be instrument specific. For a rough estimate,
we assume that for a given amount of telescope time an imaging survey can reach roughly 1 mag deeper in the continuum than spectroscopy. \citet{Lee2014a,Schmidt2019} argue that the limiting UV magnitude for spectroscopic tomography achievable with the current 8-10m class telescope is about the apparent magnitude of $m_{\rm UV}\sim25.0$ after $\sim5$ hours exposure.
With a similar exposure time, Subaru/HSC imaging can typically reach the limiting magnitude of $\sim26.0$ in a broad band \citep{Aihara2021}. For a field-of-view (FoV) of a single pointing of narrow-band and spectroscopic tomographic survey, we assume that we use Subaru/HSC for NB tomography with a FoV of $1.76\rm\,deg^2$ and Keck/DEIMOS for spectroscopic tomography with a FoV of $0.0178\,\rm deg^2$ ($=4\times16\,\rm arcmin^2$).

\begin{figure}
  \hspace{-1cm}
	\includegraphics[width=1.25\columnwidth]{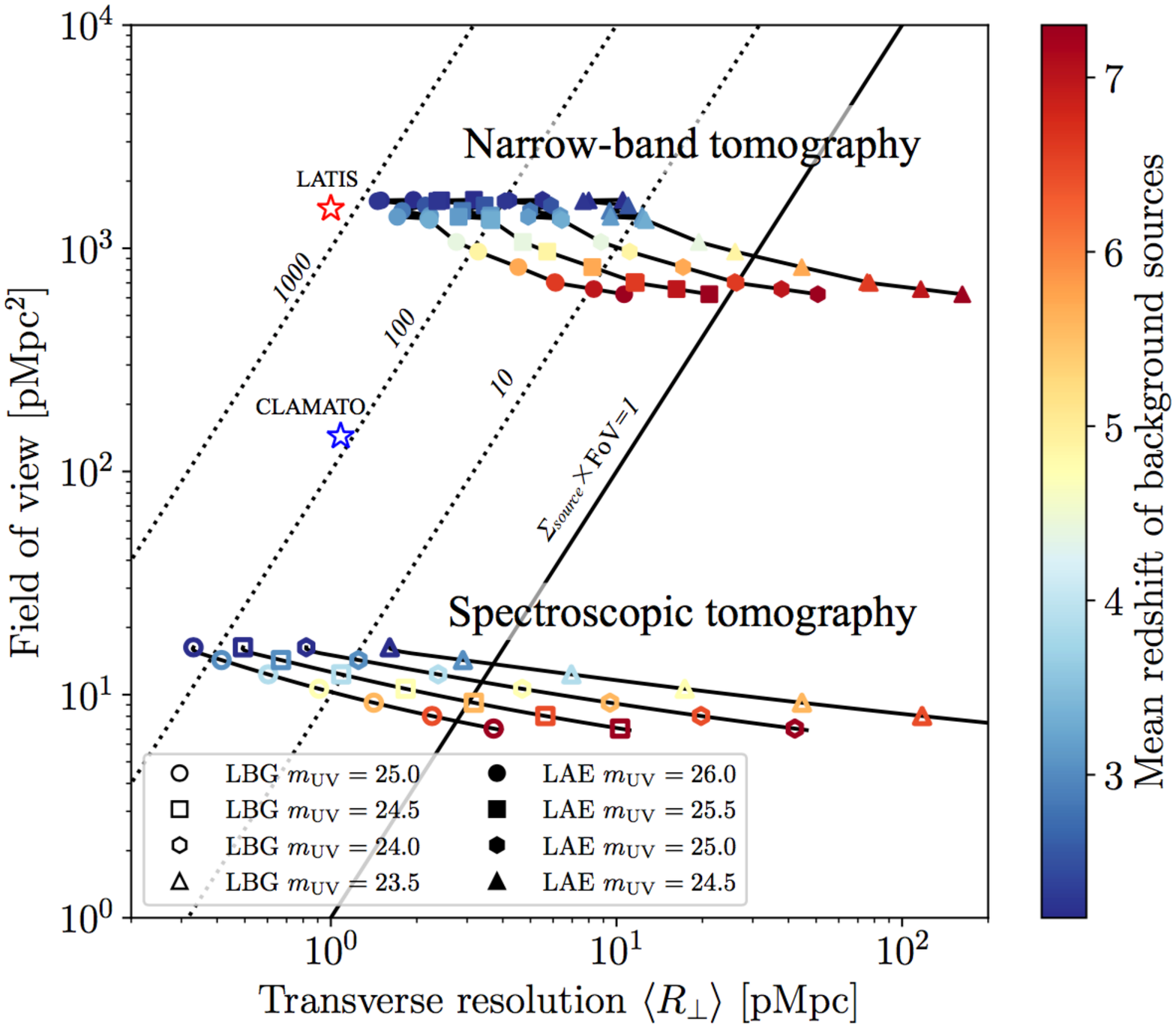}
	\vspace{-0.4cm}
\caption{Figure of merit of narrow-band and spectroscopic IGM tomography. Each line represents the field-of-view and transverse resolution of a IGM tomography at a given apparent UV magnitude depth for various mean redshift of the background galaxies.  A tomographic survey at the upper left corner has a larger field-of-view and high spatial resolution. The diagonal lines indicate the expected number of background galaxies within a field-of-view of a survey. In order for a survey to have a sensible number of background galaxies, it needs to lie at the upper left side of the solid diagonal line ($>1$ background galaxy per field of view). The figure illustrates the NB tomography has an advantage of covering a large field-of-view with a modest spatial resolution especially at a higher redshift whereas the spectroscopic tomography is suited to obtain high spatial resolution map in a small portion of the sky.}
    \label{fig:figure_of_merit}
\end{figure}

For spectroscopic tomography, the surface number density of the background galaxies is
\begin{equation}
    \Sigma_{\rm \scriptscriptstyle LBG}(<M^{\rm lim}_{\rm UV})=\int_{z_{\rm min}}^{z_{\rm max}}\left|\frac{dl_p}{dz}\right|(1+z)^3\int_{-\infty}^{\Muv^{\rm lim}}\frac{dn}{d\Muv}d\Muv,
\end{equation}
where $z_{\rm max}=(\lambda_\alpha/1040{\rm\,\mbox{\AA}})(1+z_Q)-1$ and $z_{\rm min}=(\lambda_\alpha/1180{\rm\,\mbox{\AA}})(1+z_Q)-1$ set the redshift range such that the Ly$\alpha$ forest region of a background LBG can probe the Ly$\alpha$ absorption at the quasar redshift $z_Q$ \citep{Lee2014a,Schmidt2019}.
This provides a much larger line-of-sight volume for background galaxies than NB tomography. In addition, if we assume that spectroscopic redshift can be determined by the Lyman break feature, spectroscopic tomography can provide a higher surface density for background galaxies than the NB counterpart. The NB selection is limited within the NB filter width, meaning that while the photometric background sources can be fainter, this is balanced out by a smaller search volume for the background sources. We compare the spatial resolutions and FoV of narrow-band and spectroscopic tomography at various redshifts for a single pointing
in Figure \ref{fig:figure_of_merit}. Indeed, at $z\sim2-3$ spectroscopic tomography can typically achieve a higher spatial resolution than the NB tomography because a larger line-of-sight volume is available to locate suitable background galaxies. At higher redshifts $z\sim4-6$, the NB tomography can provide a comparable surface number density, i.e. spatial resolution, of background galaxies to the spectroscopic tomography. This is because at higher redshifts, the increasing fraction of star-forming galaxies shows Ly$\alpha$ emission as they become younger and dust-free at higher redshifts. At $z\sim2-3$ only a small fraction ($X_\alpha\sim10-20\%$) of star-forming galaxies shows strong ($\rm REW>25\,\mbox{\AA}$) Ly$\alpha$ emission. This fraction increases to $X_\alpha\sim40\%$ from $z\sim2$ to $6$, making the NB tomographic technique an valuable approach over spectroscopic method at higher redshifts.

One major advantage of NB tomography
is the large increase in the field of view. Compared to existing spectroscopic tomographic surveys, CLAMATO \citep{Lee2018} and LATIS \citep{Newman2020},  NB tomography can achieve a comparable sky coverage to the multi-pointing spectroscopic tomography with a single pointing. This makes it particularly suitable to search for coherent fluctuations in Ly$\alpha$ forest absorption, e.g. by protoclusters or quasar light echoes, in a single redshift slice. NB tomography provides an efficient means to survey a large field of view and identify interesting large-scale structures in the IGM. Furthermore, as the imaging data for the NB tomographic survey naturally allows us to select background galaxies using a dropout technique, a spectroscopic follow-up campaign can boost the number of background galaxies as well as to spectroscopically detect the Ly$\alpha$ forest transmission. This enables us to potentially examine the IGM structure at improved spatial resolution and in 3D in detail. For example, NB tomography with a depth of 25 mag for the background UV continuum galaxies will provide a dropout sample suitable for spectroscopic tomography.

\section{Photometric light-echo tomography}\label{sec:mock}

Our estimate for the
surface number density of background LAEs set the general
requirements for the photometric IGM tomography to map quasar light echoes given a desired spatial/temporal resolution.
In this section, we use cosmological hydrodynamic simulations and a simple quasar emission model to construct mock observations which treat the various sources of noise in the measurement including photometric errors, systematic error in the background galaxy SED template, the Poission fluctuations in the number of background galaxies, and cosmic variance resulting from the fluctuations in the IGM.

\subsection{Cosmological hydrodynamic simulation}

We use a cosmological hydrodynamic simulation performed with the Eulerian code \textsc{NyX} \citep{Almgren2013,Lukic2015}. The simulation was performed in a large box of $100h^{-1}\rm cMpc$. This is sufficient to cover a major portion of HSC's $1.78\,\rm deg^2$ FoV corresponding to ${\rm FoV}^{1/2}\simeq90-130h^{-1}\rm cMpc$ at $z\sim2-6$. The hydrodynamics is computed on a fixed grid of $4096^3$ resolution elements and the same number of the dark matter particles, corresponding to the uniform spatial resolution of $24.4h^{-1}\rm ckpc$. This provides numerically converged Ly$\alpha$ forest statistics at the one percent level across the simulation domain suitable for examining the IGM properties in detail \citep{Lukic2015}. The simulation assumes a homogeneous optically thin UV background from \citet{Haardt2012}.

Ly$\alpha$ optical depths are computed according to \citet{Schmidt2018,Schmidt2019} using the simulated density, velocity and temperature along the skewers. The mean photoionization rate is rescaled to match the observed effective Ly$\alpha$ optical depth $\tau_{\rm eff}=0.00126\times e^{3.294\sqrt{z}}$ \citep{Onorbe2017}, but the thermal structure is kept unchanged from the original simulation output.

We include the photoionization by a foreground quasar. We assume a quasar emits isotropically and is located in a dark matter halo with mass $\gtrsim10^{12}\rm\,M_\odot$ \citep{Shen2009,White2012}.
As the simulation box is periodic, we recenter the simulation box at the position of the quasar for our convenience without loss of generality. We assume the absolute UV magnitude of $M_{\rm 1450}=-28.0$ for the quasar and use the \citet{Lusso2015} quasar
template for the spectral energy distribution with EUV slope $\alpha=-1.7$ beyond $912$\,\AA. The photoionization rate
source by the quasar for radiation emitted at time $t$ is
\begin{align}
    \Gamma_{\rm QSO}^{\rm HI}
    =\int^{\infty}_{\nu_{\rm HI}}\frac{\sigmaHI}{h\nu}\frac{L^{\rm QSO}_\nu(t)}{4\pi R^2}e^{-R/\lambda_{\rm mfp}}{\rm d}\nu
\end{align}
where $L_\nu(t)$ is the specific luminosity of the quasar lightcurve, $R$ is the proper 3D distance from the quasar, $\lambda_{\rm mfp}$ is the mean free path of the ionizing photons, and $\sigmaHI\propto(\nu/\nu_{\rm HI})^{-3}$ is the $\HI$ photoionization cross-section. As we work in the optically thin limit and thus ignore self-shielding by Lyman limit systems, we assume the mean free path of inifinte length $\lambda_{\rm mfp}=\infty$. \citet{Worseck2014} reports the measured value of the mean free path of $\lambda_{\rm mfp}=22.2\pm2.3\,\rm pMpc$ at $z=4.56$, which is larger than the expected size of the quasar light echo and the HSC's FoV. Thus, assuming $\lambda_{\rm mfp}=\infty$ is appropriate for the scales and redshifts of our interest.

In order to simulate the quasar light-echo effect, we need to set a model for quasar lightcurve. For simplicity we assume a lightbulb model with a quasar age of $t_{\rm age}$. At each location of the IGM separated by proper transverse $R_{\perp}$ and line-of-sight $R_\parallel$ distance away from the quasar has quasar photoionization rate according to equation (\ref{eq:quasar_photoionization}). In practice, a more accurate scheme is used
including the effect of cosmic expansion used to simulate the quasar light-echoes according to \citet{Schmidt2019}.
Note that the lightbulb model is clearly an oversimplification of a more realistic variable quasar lightcurve, which may occur as a result of merger or feedback-regulated driven mechanism of gas feeding onto the circumnuclear region, and/or episodic super-Eddington accretion phases of the accretion disk. However, it provides a simple well-defined measure of the characteristic timescale for the quasar activity as $t_{\rm age}$. Thus, this serves as a useful effective parameter to assess the observational requirements to constrain the quasar-active growth history of a SMBH from light-echo tomography.

\subsection{Mock observations}

We generate a mock photometric sample of background LAEs in the field of the foreground quasar to forward model a NB light-echo tomographic survey. In this procedure, we first randomly choose $N$ number of background LAEs assuming the Poisson distribution,
\begin{equation}
P(N|\bar{N})=\frac{\bar{N}^Ne^{-\bar{N}}}{N!},
\end{equation}
where $\bar{N}=\Sigma_{\rm LAE}(<m_{\rm uv}^{\rm lim})\times {\rm FoV}$ is the mean number of background LAEs above the limiting UV magnitude of the survey. We use the empircial model for the surface number density of LAEs
as discussed in Section \ref{sec:background_sources}. We then distribute the $N$ background LAEs at random transverse positions $\{\boldsymbol{r}_\perp\}_{i=1\dots,N}$ within the FoV assuming a uniform random distribution. Since the physical distance between the foreground quasar and background LAEs is large, we expect no spatial correlation with the location of foreground quasar. We ignore the effect of background LAE clustering, which may produce a clustered sampling of Ly$\alpha$ forest sightlines. While this leads to a different window function for IGM tomography, the effect should not significantly
modify our result.

For each background LAE, we assume the intrinsic spectrum to follow a power-law SED $f_\nu=f_{1500}(\Muv,z)(\nu/\nu_{1500})^{-(\beta+2)}$ in the rest-frame wavelength range between $1026$\,\AA~and $2000$\,\AA, which is characterised by the normalization $f_{1500}(\Muv,z)$ at 1500\,\AA~and the continuum slope $\beta$. This is a good approximation for a galaxy with little dust extinction whose UV continuum is dominated by the stellar continuum. The power-law spectrum is consistent with the results from stellar population synthesis models \citep[e.g. BPASS,][]{Eldridge2017} in the UV wavelength range. At a given redshift for the background LAEs,  the UV magnitude of each galaxy is drawn randomly from the UV luminosity function of LAEs (equation \ref{eq:UV_LF_LAE}) above the limiting magnitude of the survey. The value of the continuum slope $\beta$ is also randomly drawn assuming a Gaussian distribution of $\langle\beta\rangle=-1.8$ and the standard deviation of $\sigma_\beta=0.68$.
These values are determined from the best-fit Gaussian to the distribution of the $\beta$ slopes measured from the \citet{Bouwens2014} sample at $z=4-6$.

Along each transverse coordinate $\boldsymbol{r}_\perp$ of a background LAE, we draw a skewer of Ly$\alpha$ forest transmission $e^{-\tau_\alpha}$ along the line-of-sight using the cosmological hydrodynamic simulation. Using the intrinsic galaxy spectrum of the LAE and the simulated Ly$\alpha$ forest transmission, we compute the NB photometric flux of the background LAE, which measures the transmitted Ly$\alpha$ forest flux along the background LAEs at the redshift of the foreground quasar,
\begin{equation}
f_{\rm NB}=\frac{\int e^{-\tau_\alpha} f_\nu  T_{\rm NB}(\nu)d\nu}{\int T_{\rm NB}(\nu)d\nu}\approx T_{\rm IGM}f_{\rm NB}^{\rm intr},
\end{equation}
where
\begin{equation}
T_{\rm IGM}=\frac{\int e^{-\tau_\alpha}  T_{\rm NB}(\nu)d\nu}{\int T_{\rm NB}(\nu)d\nu}
\end{equation}
is the NB-averaged Ly$\alpha$ forest transmission and $T_{\rm NB}(\nu)$ is the filter transmission curve of the foreground NB filter. We assume the central wavelength of the NB filter is exactly matched to the redshift of the foreground quasar. $f_{\rm NB}^{\rm intr}$ denotes the NB-integrated intrinsic flux $f_{\rm NB}^{\rm intr}\equiv\int f_\nu~T_{\rm NB}(\nu)d\nu/\int  T_{\rm NB}(\nu)d\nu$. We use the realistic filter transmission curve for Subaru/HSC including CCD quantum efficiency and transmittance through the dewar window and the primary focus unit.

We add photometric noise to simulate the observed NB flux as
\begin{equation}
f_{\rm NB}^{\rm obs}=f_{\rm NB}+\delta f_{\rm NB},
\end{equation}
where $\delta f_{\rm NB}$ is the observational noise of the NB photometry.
We assume random Gaussian (background-limited) noise for $\delta f_{\rm NB}$ with the rms level $\sigma_{\rm NB}=10^{-(m_{\rm NB}^{\rm lim}+48.59)/2.5}/{\rm SNR}_{\rm NB}$, which is determined by the limiting NB magnitude $m_{\rm NB}^{\rm lim}$ at a signal-to-noise ratio $\rm SNR_{\rm NB}$.

Similarly for each background LAE, we model the observed BB flux covering redward
of Ly$\alpha$ emission as
\begin{equation}
    f_{\rm BB}^{\rm obs}=f_{\rm BB}+\delta f_{\rm BB},~~~~~
    f_{\rm BB}=\frac{\int f_\nu  T_{\rm BB}(\nu)d\nu}{\int T_{\rm BB}(\nu)d\nu},
\end{equation}
where $\delta f_{\rm BB}$ is the BB noise at a limiting BB magnitude $m_{\rm BB}$ with
signal-to-noise of ${\rm SNR}_{\rm BB}$, which is computed using the same procedure as the NB filter.

This procedure gives a mock sample of $N$ background LAEs with observed NB and BB fluxes at random transverse positions. The simulated mock thus consists of $\{f_{\rm NB}^{\rm obs},f_{\rm BB}^{\rm obs},\boldsymbol{r}_\perp\}_{i=1,\dots,N}$ for each model of lifetime $t_Q$ and UV magnitude $M_{1450}$ of a quasar. This closely mimics the realistic photometric dataset from a NB tomographic survey targeting a quasar field.

\subsection{Measurement}

To estimate the Ly$\alpha$ forest transmission from a photometric sample, we need to infer the intrinsic NB flux without the influence of Ly$\alpha$ forest transmission using the information from observed BB fluxes redward of Ly$\alpha$ line. When only a single BB filter is available,
we assume a template galaxy (power-law) spectrum $f_\nu^{\rm temp}=f_{1500}^{\rm temp}(\nu/\nu_{1500})^{-(\beta_{\rm temp}+2)}$ with an assumed value of continuum slope $\beta_{\rm temp}$. Since we do not know a priori the continuum slope of background LAEs, this need not be the same as our input $\beta$ slope used while generating the mock observation. We then fit the template spectrum to the observed BB flux to determine the normalisation $f_{1500}^{\rm temp}$. Once the best-fit intrinsic galaxy spectrum is determined, we can estimate the intrinsic NB flux as
\begin{equation}
    f_{\rm NB}^{\rm intr}=\frac{\int f_\nu^{\rm temp}  T_{\rm NB}(\nu)d\nu}{\int T_{\rm NB}(\nu)d\nu}.
\end{equation}
Thus the measured Ly$\alpha$ forest transmission $\hat{T}_{\rm IGM}$ along each background LAE is given by the ratio between the observed and inferred intrinsic NB fluxes,
\begin{equation}
    \hat{T}_{\rm IGM}=\frac{f_{\rm NB}^{\rm obs}}{f_{\rm NB}^{\rm intr}}.
\end{equation}
Note that this is a noisy estimate of the true underlying value of the Ly$\alpha$ forest transmission $T_{\rm IGM}$, which we compute from the hydrodynamic simulation for our mock survey.
This estimated $\hat{T}_{\rm IGM}$ is is affected by the photometric noises in both NB and BB filters and the systematic error from the difference between the true intrinsic galaxy spectrum and the assumed template. In the following part of Section \ref{sec:mock}, we adopt this procedure for estimating $\hat{T}_{\rm IGM}$, which is informative to examine the various sources of errors on the 2D tomographic map of a quasar light echo.

If multiple BB fluxes are available, one can better constrain the intrinsic galaxy spectrum by simultaneously fitting both for the normalisation and continuum slope. In fact, one can recast the whole procedure of estimating Ly$\alpha$ forest transmission as a single SED fitting procedure to simultaneously estimate Ly$\alpha$ forest transmission $T_{\rm IGM}$, UV continuum slope $\beta$, and the normalisation $f_{1500}^{\rm temp}$. This provides a natural framework to propagate both observational photometric noise and systematic error in the assumed intrinsic galaxy spectrum to the final measurement of Ly$\alpha$ forest transmission. Indeed, as we will introduce in Section \ref{sec:inference}, our statistical inference framework to constrain the quasar lifetime from NB IGM tomography is based on this approach.

\begin{figure*}
	\includegraphics[width=\textwidth]{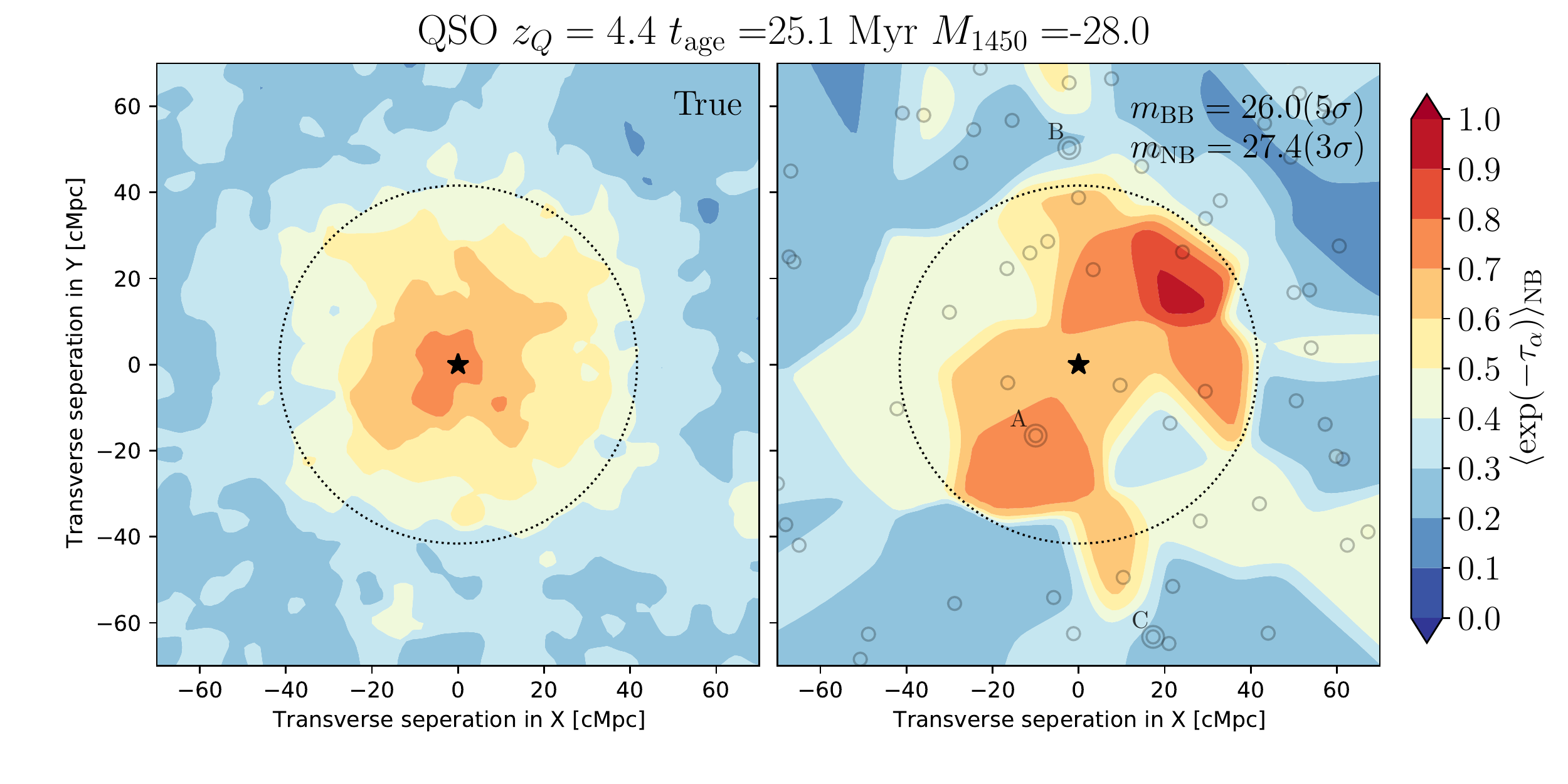}
	\hspace*{0.55cm}
	\includegraphics[width=\textwidth]{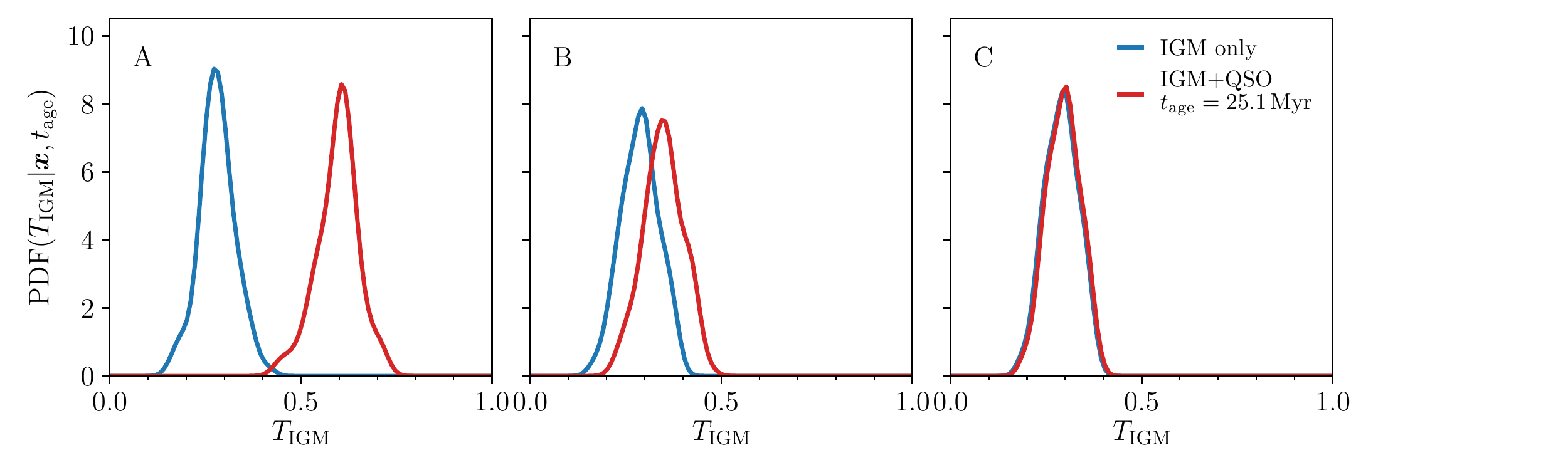}
	\vspace{-0.5cm}
\caption{({\bf Top panels}) Example of the reconstructed 2D $\HI$ Ly$\alpha$ forest transmission map around a quasar with $t_{\rm age}=25.1\rm\,Myr$ and $M_{1450}=-28.0$ at $z=4.4$ via NB tomography with the limiting BB magnitude $m_{\rm BB}=26.0\,(5\sigma)$ and the limiting foreground NB magnitude $m_{\rm BB}=27.4\,(3\sigma)$. The left panel show the 2D reconstructed map with a smoothing length of $2\rm\,cMpc$ for noiseless photometry and a large number of background galaxies. The right panel show the 2D reconstructed map with a smoothing length of $5\rm\,cMpc$ for a mock survey including the effects of photometric noise, continuum error of the background LAEs, and finite Poisson sampling. The location of background LAEs are indicated by solid circles and the location of the quasar is indicated by star symbol. The dotted circle indicates the region of quasar influence $R=ct_{\rm age}$. ({\bf Bottom panels}) The probability distribution function of $T_{\rm IGM}$, ${\rm PDF}(T_{\rm IGM}|\boldsymbol{r}_\perp,t_{\rm age})$, at the indicated  sightlines A, B, C. This is computed using all random realizations of the quasar-host halo in the simulation at the fixed sightline locations relative to the quasar in order to evaluate the impact of fluctuating IGM densities. The distributions of $T_{\rm IGM}$ with and without the quasar transverse proximity effect (red: IGM+QSO, blue: IGM only) at $t_{\rm age}=25.1\,\rm Myr$ are shown.}
    \label{fig:2D_map}
\end{figure*}

\subsection{2D tomographic map reconstruction}

The dense sample of background LAEs permits the reconstruction of the 2D map of the NB-averaged Ly$\alpha$ forest transmission around the foreground quasar.  In order to create a 2D reconstructed map, we gaussian interpolate a set of estimated Ly$\alpha$ forest transmissions $\hat{T}_{{\rm IGM},i}$ measured at observed transverse coordinates $\boldsymbol{r}_{\perp,i}$ along the background LAEs $i=1,\dots,N$. More specifically, we use a simple non-parametric regression method called the Nadaraya-Watson estimator,
\begin{equation}
    \hat{T}_{\rm IGM}^{\rm 2D}(\boldsymbol{r}_\perp)=\frac{\sum_{i=1}^{N} \hat{T}_{{\rm IGM},i}K_\sigma(\boldsymbol{r}_\perp-\boldsymbol{r}_{\perp,i})}{\sum_{i=1}^{N} K_\sigma(\boldsymbol{r}_{\perp}-\boldsymbol{r}_{\perp,i})}
\end{equation}
where $K_\sigma(\boldsymbol{r}_\perp-\boldsymbol{r}_{\perp,i})=(2\pi \sigma^2)^{-1}\exp[-|\boldsymbol{r}_\perp-\boldsymbol{r}_{\perp,i}|^2/(2\sigma^2)]$ is the 2D Gaussian kernel with the smoothing length (standard deviation) of $\sigma$. We choose the smoothing length based on the expected typical sightline separation of background LAEs at a given limiting UV magnitude of a survey as the full-width-half-maximum, $\sigma=\Sigma_{\rm LAE}^{-1/2}(<m_{\rm uv}^{\rm lim})/(2\sqrt{2\ln2})$. More sophisticated methods such as Wiener filtering \citep{Pichon2001,Caucci2008,Lee2014a,Lee2014b}, local polynomial estimator \citep{Cisewski2014}, and optimisation techniques \citep{Horowitz2019,Li2021} were examined and applied for the 3D reconstruction for spectroscopic tomographic maps in the past. Here we only use the 2D tomographic map for visualisation purpose, and do not use it
for our statistical inference framework. Thus, the simple estimator suffices for the scope of this paper.

To demonstrate the reconstruction of 2D tomographic map from our mock survey, we assume a survey targetting LAEs background of a $z=4.4$ quasar with a limiting BB and foreground NB magnitudes of $m_{\rm BB}^{\rm lim}=26.0$ at $5\sigma$ and $m_{\rm BB}^{\rm lim}=27.4$ at $3\sigma$. For $z=4.4$ quasar light-echo tomography with Subaru/HSC, this corresponds to $i2$ for BB and NB656 for foreground NB with the $z=4.9$ background LAEs located by NB718. We also assume an isotropically emitting quasar model at $z_Q=4.4$ with the lifetime of $t_{\rm age}=25.1\,\rm Myr$ and UV magnitude $M_{1450}=-28.0$. This ultra-luminous quasar luminosity which can be selected from e.g. SDSS DR16 and/or Pan-STARRS quasar catalogues \citep{Schindler2019,Lyke2020} is ideal for the light echo tomographic experiment because it is expected to cause the strongest enhancement in Ly$\alpha$ forest transmission and have the largest proximity zone \citep{Schmidt2019}.
Figure~\ref{fig:2D_map} shows an
example of the 2D reconstructed Ly$\alpha$ forest transmission map around the quasar from our mock survey.
As it will be discussed in Section \ref{sec:depth}, this corresponds to approximately a total of $19.6$ hours of exposure time for a minimal set of NB656, NB718, $r$, and $i$ imaging required for the NB IGM tomographic survey.
It shows that the reconstructed 2D tomographic map is the sparsely sampled interpolated version of the true map in the limit of infinite signal-to-noise and the infinite number of background galaxies. Despite the observational limitations including photometric noise, UV continuum uncertainty in $\beta$ slope, and the finite Poisson sampling of background LAEs, the ionizing light-echo from the quasar is clearly visible as a transverse proximity effect in the reconstructed Ly$\alpha$ forest transmission map. The transverse proximity zone extends out to the radius of the delay time surface $R=c t_{\rm age}$ corresponding to $t_{\rm age}=25.1\rm\,Myr$ as indicated by the dotted circle. Inside the transverse proximity zone (e.g. location A in Figure \ref{fig:2D_map}), the Ly$\alpha$ forest transmission is clearly enhanced larger than $5\sigma$ density fluctuations of the IGM.

\subsection{Observational requirement and errors}

When searching for the quasar light-echo signal in the NB tomography, there are three other sources of fluctuations in the observed Ly$\alpha$ forest transmission apart from the quasar photoionization; (1) photometric noise, (2) density fluctuations from the IGM, (3) the uncertain UV continuum of the background LAEs. We examine the impact of each source of uncertainty on NB tomography with the aim of understanding the requirements to overcome these sources of noise.

\begin{figure*}
	\includegraphics[width=\textwidth]{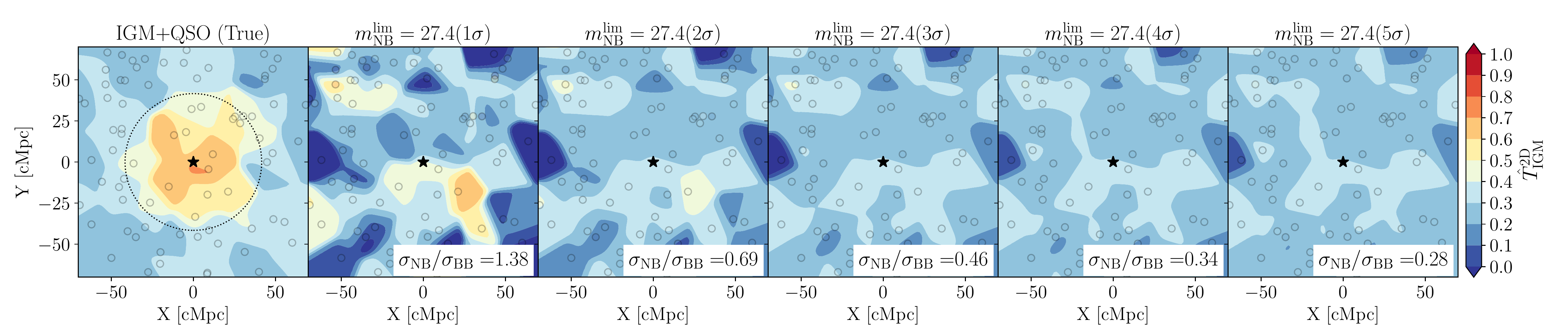}
	\vspace{-0.3cm}
\caption{The effect of photometric noise on the reconstructed 2D tomographic map with the smoothing length of $5\rm \,cMpc$. We compare the true model at $z=4.4$ QSO with $M_{1450}=-28.0$ and $t_{\rm age}=25\rm\,Myr$ including the IGM fluctuations but without photmetric noise and continuum uncertainty. The various SNR of the foregorund NB filter shows the fluctuations of photometric noise around the fixed mean Ly$\alpha$ forest transmission $T_{\rm IGM}=0.28$ at $z=4.4$. The figure shows the $\rm SNR_{NB}\gtrsim3$ is sufficient to avoid the fictitious photometric noise fluctuations mimicking the quasar's light echo signal (transverse proximity effect).}
    \label{fig:error_noise_map}
\end{figure*}

\subsubsection{Photometric noise}

Once the required limiting BB depth $m_{\rm BB}^{\rm lim}$ is determined based on the desired surface number density of background LAEs, the key remaining observational requirement is the depth in the foreground NB filter. The NB imaging needs to be deep enough $m_{\rm NB}^{\rm lim}>m_{\rm BB}^{\rm lim}$ (more precisely $\sigma_{\rm NB}<\sigma_{\rm BB}$) so that one can place a meaningful constraint ($\hat{T}_{\rm IGM}<1$) on the Ly$\alpha$ forest transmission  around the foreground quasar. In order to measure the Ly$\alpha$ forest transmission $T_{\rm IGM}$ along a background LAE with a limiting BB magnitude $m_{\rm BB}^{\rm lim}$ which covers the UV continuum, the limiting NB magnitude needs to be
\begin{equation}
m_{\rm NB}^{\rm lim}=m_{\rm BB}^{\rm lim}-2.5\log_{10}T_{\rm IGM}\left(\frac{\lambda_{\rm BB}}{\lambda_{\rm NB}}\right)^{-(\beta+2)},\label{eq:limmag}
\end{equation}
where the factor of $(\lambda_{\rm BB}/\lambda_{\rm NB})^{-(\beta+2)}$ comes from the required extrapolation of observed BB flux to estimate the intrinsic NB flux for a galaxy with UV continuum slope $\beta$ ($\lambda_{\rm BB}$ and $\lambda_{\rm NB}$ are the central wavelengths of the BB and foreground NB filter). The most conservative choice is to require $T_{\rm IGM}=e^{-\tau_{\rm eff}(z)}$ so that we detect the mean Ly$\alpha$ forest transmission at a redshift of the foreground NB filter. This guarantees the direct detection of mean Ly$\alpha$ forest transmission along individual background LAEs and ensures that the contrast between the mean and enhanced Ly$\alpha$ forest transmissions by the quasar light-echoes is detected. Based on this consideration, assuming $e^{-\tau_{\rm eff}}(z)\simeq0.28$ and $\beta=-1.8$ for $z=4.4$ NB tomography, the limiting NB magnitude needs to be deeper by
\begin{equation}
    m_{\rm NB}^{\rm lim}-m_{\rm BB}^{\rm lim}\simeq1.4,
\end{equation}
compared to the limiting BB magnitude.

This sets quite a stringent requirement on the foreground NB depth if we require a $5\sigma$ detection of mean Ly$\alpha$ forest transmission along the individual LAEs, e.g. $m_{\rm NB}^{\rm lim}=27.4\,(5\sigma)$ for $m_{\rm BB}^{\rm lim}=26.0\,(5\sigma)$. However, because our goal is a statistical detection of the quasar light-echo signal in the tomography, we do not necessarily need a $5\sigma$ detection of NB flux for each background LAE. We can statistically average over the NB fluxes along many background LAE sightlines. Thus the signal-to-noise ratio of the limiting magnitude for the foreground NB filter can be small than 5.
Indeed, we can set a minimal requirement on the NB limiting magnitude such that
it is smaller than the typical enhanced fluctuations of the Ly$\alpha$ forest transmission by the quasar light echoes.

\begin{figure}
	\includegraphics[width=\columnwidth]{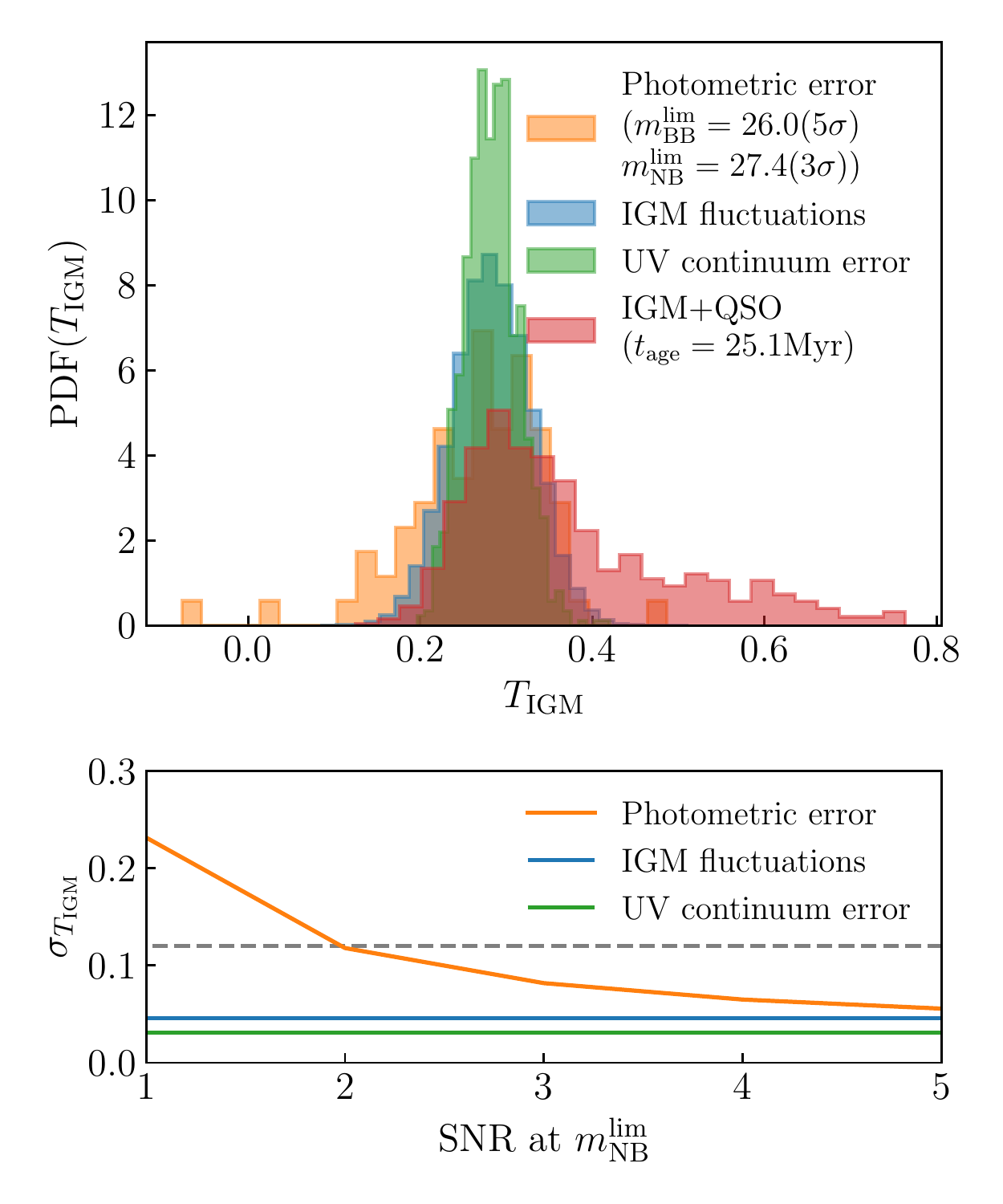}
	\vspace{-0.5cm}
\caption{The various sources of the fluctuations of NB integrated Ly$\alpha$ forest transmission $T_{\rm IGM}$ including photometric error with $m_{\rm BB}=26.0(5\sigma)$ and $m_{\rm BB}=27.4(3\sigma)$ (orange), IGM density fluctuations (blue), UV continuum uncertainty (green), QSO transverse proximity effect at $t_{\rm age}=25.1\rm\,Myr$ (red). The bottom panel compares the fluctuations due to the photometric error as a function of SNR of the NB filter with the IGM fluctuations and UV continuum uncertainty. The region above the dashed horizontal line show the fluctuations due to the quasar transverse proximity effect ($T_{\rm IGM}>0.4$). The lower photometric fluctuations below this dashed line illustrates a sufficient depth to detect the quasar light echo in the NB IGM tomography.}
    \label{fig:error_noise_histogram}
\end{figure}

Figure~\ref{fig:error_noise_map} illustrates the effect of photometric noise on the tomographic map with varying signal-to-noise ratios of the NB imaging. At low signal-to-noise ratios, one can see the fictitious transparent region due to the large photometric noise. The estimated Ly$\alpha$ forest transmissions take values outside the physically meaningful range of $0\le\hat{T}_{\rm IGM}\le1$. Increasing the signal-to-noise ratio minimises the effect of fictitious fluctuations in the Ly$\alpha$ forest transmission.
More quantitatively, Figure \ref{fig:error_noise_histogram} compares the fluctuations in $T_{\rm IGM}$ caused by the quasar light echo (red) with those by the photometric noise (orange) as well as by other sources of uncertainties (IGM fluctuations: blue, UV continuum error: green). The quasar light echoes produce a tail of highly transmissive regions of the IGM above $T_{\rm IGM}>0.4$. When the signal-to-noise ratio of the foreground NB imaging is small ${\rm SNR}_{\rm NB}\lesssim2$, the dominant source of fluctuations in the IGM tomographic map is the photometric noise. However, already at $\rm SNR_{\rm NB}\gtrsim3$, the quasar light echo signal becomes a dominant source of fluctuations. This means that a modest signal-to-noise ratio of the NB imaging is sufficient to detect the quasar light echo with NB IGM tomographic method. The effect of photometric fluctuations are limited to $\sigma_{T_{\rm IGM}}\lesssim0.10$ at $\rm SNR_{\rm NB}\gtrsim3-5$ at the NB limiting magnitude and $5\sigma$ for the BB limiting magnitude. While the photometric fluctuations are typically larger than the IGM fluctuations, the quasar light echo imprints high Ly$\alpha$ forest transmission regions of the IGM with $T_{\rm IGM}>0.4$, corresponding to $\sigma_{T_{\rm IGM}}>0.12$ relative to the mean IGM transmission. Thus for light-echo tomographic experiment, the mininal requirement is set by detecting the enhanced Ly$\alpha$ forest transmission by the quasar light echo. This relaxes the required depth of the NB imaging. As photometric noise becomes sub-dominant, we choose ${\rm SNR}_{\rm NB}=3$ at the NB limiting magnitude for our fidicual value.

\subsubsection{Density fluctuations of the IGM}

As we are interested in the transverse proximity effect due to the quasar light-echo signal, the density fluctuations of the IGM acts as a source of uncertainty in the tomographic mapping of the quasar transverse proximity effect. In Figures \ref{fig:error_noise_histogram} and \ref{fig:error_igm+cont} we quantify the fluctuations in the Ly$\alpha$ forest transmission $T_{\rm IGM}$ due to the IGM density fluctuations and compare them with those due to the quasar light echoes. The IGM density fluctuations cause a nearly Gaussian fluctuations in $T_{\rm IGM}$ with the standard deviation of $\sigma_{T_{\rm IGM}}\simeq0.05$ corresponding to $\sigma_{T_{\rm IGM}}/\bar{T}_{\rm IGM}\sim17\,\%$ relative to the mean IGM transmission at $z=4.4$. The quasar's photoionization produces a large coherent transmission within the width of NB filter ($\simeq56\rm\,cMpc$ for NB656 filter), whereas the IGM fluctuations tend to be averaged out within the length of NB filter owing to the short correlation length.  The quasar light echoes produce a regions with $T_{\rm IGM}\gtrsim0.4$. This is $\gtrsim3\sigma$ away from the IGM fluctuations. Thus, it is unlikely to confuse the IGM fluctuations with quasar light echoes.

Note that the effect of the gas overdensity around a quasar is small. In the simulation we selected a region with central quasar-host halo of mass $M_h>10^{12}\rm M_\odot$. While they reside in biased region, the quasar - gas density correlation length is small. Therefore the effect of gas overdensity around the host halo is averaged out within the NB filter width unless it resides in a rare extreme protocluster region.

\subsubsection{UV continuum slope}

\begin{figure}
	\includegraphics[width=\columnwidth]{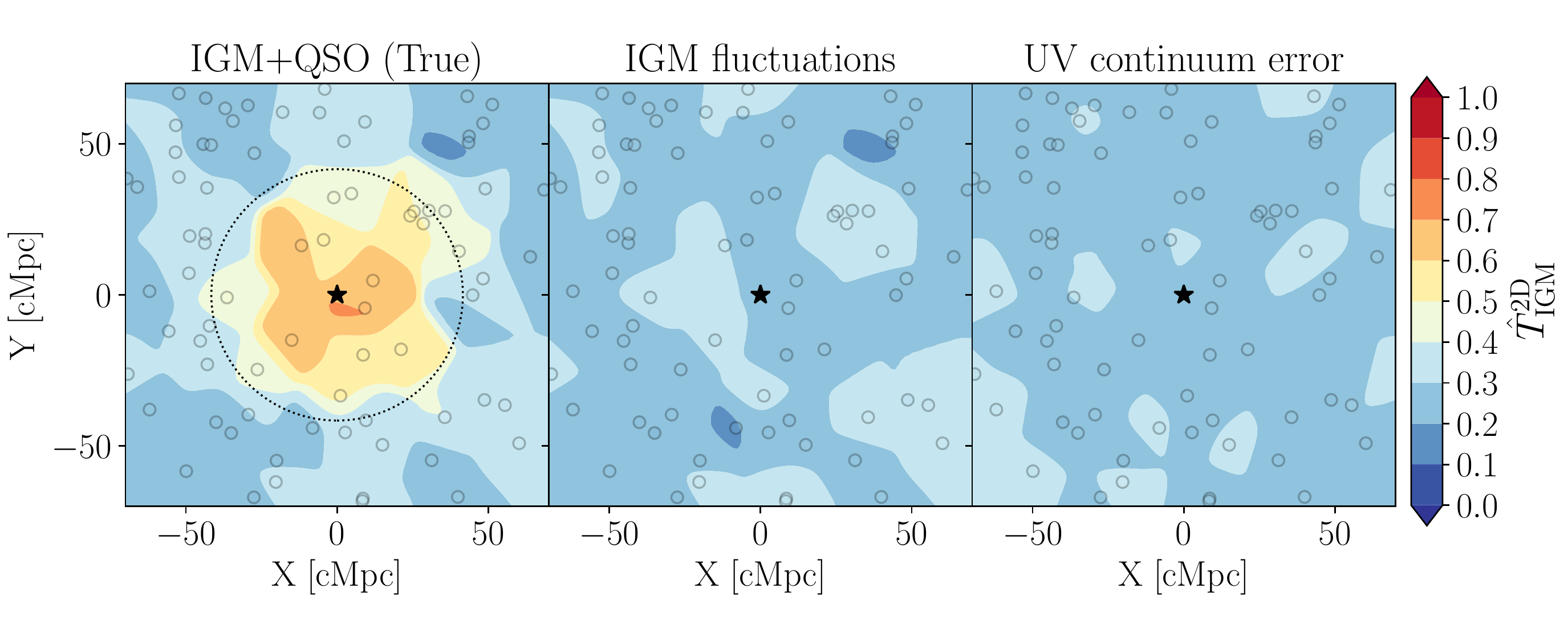}
	\vspace{-0.5cm}
\caption{The effect of IGM fluctuations and UV continuum uncertainty on the 2D tomographic map with the smoothing length of $5\,\rm cMpc$. We compare the true model (left) same as Figure \ref{fig:error_noise_map} with the maps only including the IGM fluctuations (middle) and the UV continuum uncertainty with $\beta=-1.8$ and $\sigma_\beta=0.68$. The UV continuum error map indicates the fluctuations around a fixed mean Ly$\alpha$ forest transmission $T_{\rm IGM}=0.28$ assuming the template UV continuum slope of $\beta_{\rm temp}=-1.8$.}\label{fig:error_igm+cont}
\end{figure}

The uncertainty in the UV continuum slope $\beta$ in the background galaxies introduces fictitious fluctuations in the estimated Ly$\alpha$ forest transmission map. Figure \ref{fig:error_noise_histogram} and \ref{fig:error_igm+cont} show that the UV continuum uncertainty is the sub-dominant source of error for the NB light echo tomography. Using the observed distribution of the $\beta$ slope from \citet{Bouwens2014}, the error around the mean IGM transmission is $\sigma_{T_{\rm IGM}}\simeq0.03$.
This is smaller than the IGM fluctuations and the photometric noise.

The small fluctuations of the UV continuum uncertainty are somewhat surprising. To understand this, we consider the error associated to the uncertain $\beta$ slopes in the limit of infinite signal-to-noise for NB and BB filters. The error in the template galaxy spectrum enters as a multiplicative noise to the estimated Ly$\alpha$ forest transmission. We find that the multiplicative noise is approximately given by
\begin{equation}
   \hat{T}_{\rm IGM}\approx\varepsilon_\beta T_{\rm IGM},~\varepsilon_\beta\approx\left(\frac{\lambda_{\rm BB}}{\lambda_{\rm NB}}\right)^{\beta_{\rm temp}-\beta}
\end{equation}
For $\sigma_\beta=0.68$ based on the \citet{Bouwens2014} measurement, this estimates approximately $\sim10\%$ fluctuations on $T_{\rm IGM}$. This small UV continuum slope uncertainty arises
because for $z=4.4$ NB tomographic setup with NB656 ($\lambda_{\rm NB}=6570\,$\AA) and $i2$ ($\lambda_{\rm BB}=7998\,$\AA) filters, the small wavelength separation ensures that only small extrapolation of UV continuum to Ly$\alpha$ forest region is needed to minimize the uncertainty due to the unknown galaxy SED shape.

An incorrect assumption for $\beta$ slope
introduces a slight bias in the estimated IGM transmission by a factor of $\left(\lambda_{\rm BB}/\lambda_{\rm NB}\right)^{\beta_{\rm temp}-\beta}$. Since the NB and BB filters for the $z=4.4$ NB tomography are placed close in wavelength, the bias is a negligible contribution to the total error budget. Furthermore, $\beta_{\rm temp}$ can be corrected posteriori to match the $\bar{T}_{\rm IGM}$ at large distance from the central quasar to the known $\tau_{\rm eff}$ or using a reference tomographic observation in a blank field. Thus, UV continuum uncertainty results in a nearly negligible source of noise in our tomographic map.

\begin{figure*}
	\includegraphics[width=0.95\textwidth]{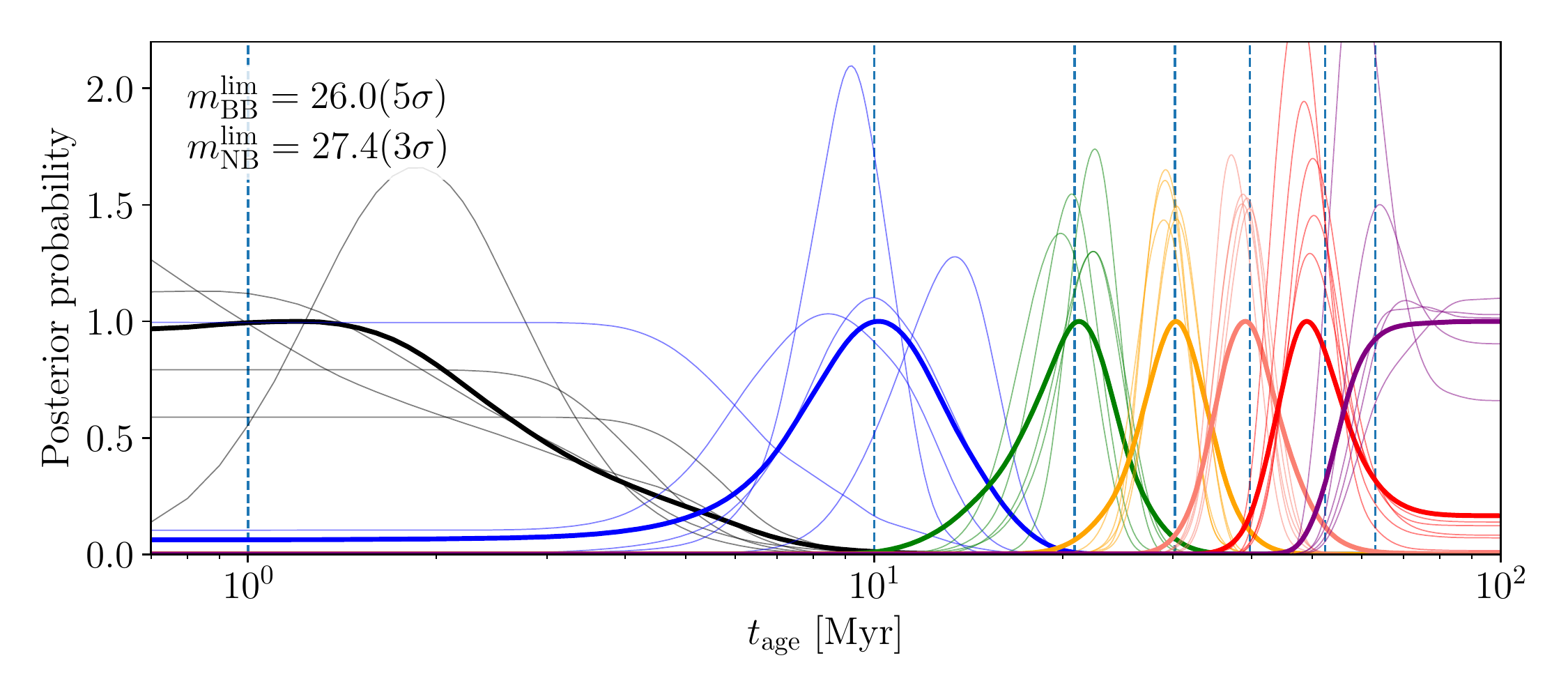}
	\vspace{-0.5cm}
\caption{Posterior probability of quasar lifetime $t_{\rm age}$ for NB tomographic survey around a $z=4.4$ quasar with limiting BB magnitude of $m_{\rm BB}^{\rm lim}=26.0(5\sigma)$ and foreground NB magnitude of $m_{\rm BB}^{\rm lim}=27.4(3\sigma)$. Thick lines represent the average posterior probabilities of 25 mock realizations, and thin lines show the posterior probabilities of random 5 mock realization for varying quasar lifetime $t_{\rm age}\simeq1, 10, 20, 30, 40, 50, 60$ Myr (black, blue, green, yellow, salmon, red, purple from left to right). The input quasar lifetimes shown by vertical dashed lines. For visualization purpose, at each quasar lifetime we scaled the posteriors with the maximum value of the average posterior probability.}
    \label{fig:posterior}
\end{figure*}

\section{Statistical inference}\label{sec:inference}

\subsection{Bayesian inference framework}

A visual inspection of the reconstructed 2D Ly$\alpha$ forest map as shown in Figure \ref{fig:2D_map} gives a first impression of the quality of NB IGM tomography of quasar light echoes.
In this section, we quantify the ability of NB tomography to constrain the properties of the quasar lightcurve.
Here we introduce a Bayesian inference framework to examine how well a NB tomographic survey constrains the quasar lifetime. For simplicity, we use a quasar lightbulb model where the quasar is turned on with the UV magnitude $M_{1450}$ for a duration of $t_{\rm age}$. As mentioned above, while this is a simplification, this acts as a useful figure-of-merit to quantify the constraining power of the NB tomography. We will examine the application of the NB tomography for more complex lightcurves in the future work.

The NB tomography delivers data comprising
a set of observed NB and BB fluxes of background LAEs. The probability that one can observe a NB flux $f_{\rm NB}^{\rm obs}$ and BB flux $f_{\rm BB}^{\rm obs}$ for a $i$-th background LAE at a given projected location $\boldsymbol r_\perp$
relative to the foreground quasar with lifetime $t_{\rm age}$ under the observational noises
on the NB and BB fluxes,  $\sigma_{\rm NB}$ and $\sigma_{\rm BB}$, can be written as
\begin{align}
&P_i(f_{\rm NB}^{\rm obs},f_{\rm BB}^{\rm obs}|t_{\rm age},\boldsymbol{r}_\perp,f_{\rm 1500},\beta,\sigma_{\rm NB},\sigma_{\rm BB})= \nonumber \\
&\mathcal{N}(f_{\rm BB}^{\rm obs}|f_{\rm BB},\sigma_{\rm BB})\int \mathcal{N}(f_{\rm NB}^{\rm obs}|f_{\rm NB},\sigma_{\rm NB})P(T_{\rm IGM}|\boldsymbol{r}_\perp,t_{\rm age})dT_{\rm IGM},
\end{align}
where $f_{\rm NB}=T_{\rm IGM}f_{\rm NB}^{\rm intr}$ and $P(T_{\rm IGM}|\boldsymbol{r}_\perp, t_{\rm age})$ is the probability distribution function (PDF) of the NB-integrated
Ly$\alpha$ forest transmission $T_{\rm IGM}$ at a given transverse coordinate $\boldsymbol{r}_\perp$ and a true quasar lifetime $t_{\rm age}$.
We forward model this PDF using the Gaussian kernel density estimation (KDE) with the width calculated by the Scott's rule
based on the 100 random realizations at each location $\boldsymbol{r}_\perp$
drawn from the cosmological hydrodynamic simulation for each quasar model with lifetime $t_{\rm age}$.
The intrinsic BB and NB fluxes, $f_{\rm BB}$ and $f_{\rm NB}^{\rm intr}$, depends on the 1500\,\AA~flux $f_{1500}$ and $\beta$ slope. The Gaussian distribution with mean $\mu$ and variance $\sigma$ is denoted by $\mathcal{N}(x|\mu,\sigma)=(2\pi\sigma^2)^{-1/2}\exp\left[-(x-\mu)^2/(2\sigma^2)\right]$.

For $N$ background LAEs, assuming all sightlines are statistically independent to each other, the likelihood of the data given a model is given by
\begin{align}
\mathcal{L}(\{f_{{\rm NB}}^{\rm obs}\, , f_{{\rm BB}}^{\rm obs}\}_{i=1,\dots,N}|t_{\rm age},\{\boldsymbol{r}_\perp,f_{{\rm 1500}},\beta\}_{i=1,\dots,N},
\sigma_{\rm NB},\sigma_{\rm BB}) \nonumber \\
~~~~~=\prod_{i=1}^{N} P_i(f_{{\rm NB}}^{\rm obs}\,, f_{{\rm BB}}^{\rm obs}|t_{\rm age},\boldsymbol{r}_{\perp},f_{{\rm 1500}},\beta,\sigma_{\rm NB},\sigma_{\rm BB}).
\end{align}
The assumption of the statistically independent sightlines are reasonable since the transverse correlation of the Ly$\alpha$ forest transmission is is expected to be smaller than the typical sightline separation of the background galaxies achievable with the NB tomography on the exiting instruments and 8-10m telescopes.

By the Bayes' theorem, we can express the posterior probablity of the quasar lifetime given the data as $\rm (posterior)\propto (prior)\times(likelihood)$.
By marginalising over the UV continuum slope and flux normalization of the background LAEs,
we find the posterior probability of the quasar lifetime as,
\begin{align}
&P(t_{\rm age}|\{f_{{\rm NB}}^{\rm obs}, f_{{\rm BB}}^{\rm obs}, \boldsymbol{r}_{\perp}\}_{i=1,\dots,N},\sigma_{\rm NB},\sigma_{\rm BB})\propto \nonumber \\
&~~~~~~~~~~~~~~~~~~~~~~~~~~~~~~~~~~~\prod_{i=1}^{N}
P_i(t_{\rm age}|f_{{\rm NB}}^{\rm obs}, f_{{\rm BB}}^{\rm obs}, \boldsymbol{r}_{\perp},\sigma_{\rm NB},\sigma_{\rm BB}),
\end{align}
where we have defined a posterior probablity of $t_{\rm age}$ from a single background LAE as,
\begin{align}
&P_i(t_{\rm age}|f_{{\rm NB}}^{\rm obs}, f_{{\rm BB}}^{\rm obs}, \boldsymbol{r}_{\perp},\sigma_{\rm NB},\sigma_{\rm BB})\propto \nonumber \\
&\int_0^\infty df_{\scriptscriptstyle 1500}P(f_{\scriptscriptstyle 1500})\int_{-\infty}^\infty d\beta P(\beta) P_i(f_{{\rm NB}}^{\rm obs}\,, f_{{\rm BB}}^{\rm obs}|t_{\rm age},\boldsymbol{r}_\perp,f_{\scriptscriptstyle 1500},\beta,\sigma_{\rm NB},\sigma_{\rm BB}).
\end{align}
We assume a flat prior $P(f_{1500})$ for $f_{1500}$ and a Gaussian prior $P(\beta)$ for $\beta$ with mean $\bar{\beta}=-1.8$ and variance $\sigma_\beta=0.68$.

This Bayesian inference framework tries to simultaneously estimate both the quasar lifetime (via its impact on the Ly$\alpha$ forest transmission)
and the intrinsic continuum level of background LAEs using the available NB and BB flux measurements, which results in a constraint on the quasar lifetime after marginalizing over the intrinsic galaxy properties (i.e. UV continuum level $f_{1500}$ and slope $\beta$). This framework permits a straightforward generalisation for multiple BB filters to reduce the uncertanties in the intrinsic galaxy spectrum (Appendix \ref{app:multi-band}). Furthermore, this inference framework directly operates on the fluxes at the observed background LAEs instead of summary statistics such as azimuthally-averaged Ly$\alpha$ forest transmission around a quasar or on reconstructed 2D Ly$\alpha$ forest map. This allows us to utilize the full information, and avoid interpreting the processed data, which may lead to a loss of information or possible artificial correlations in the data.

\subsection{Constraint on quasar lifetime}

We evaluate the posterior constraints on the quasar lifetime using the mock realisations of NB tomography with varying survey depths. Figure \ref{fig:posterior} shows the posterior probability of quasar lifetime $t_{\rm age}$ from several mock realisations of a photometric survey with the depth of $m_{\rm NB}=27.4$ ($5\sigma$) in NB656 and $m_{\rm BB}=26.0$ ($5\sigma$) in $i2$ targetting a $z=4.4$ quasar.
It shows that our Bayesian framework successfully recovers the quasar lifetime within 68\,\% confidence interval.

The accuracy on the lifetime constraint corresponds to the typical inter-sightline separation of the background galaxies. The typical width of the posteriors matches with the temporal resolution of $\Delta t\approx10\,\rm Myr$ at $m_{\rm BB}^{\rm lim}=26.0$, consistent with equation (\ref{eq:time_resol}). At small quasar lifetime, the probability of finding a background galaxy within the quasar transverse proximity zone is small. Therefore, for $t_{\rm age}<\frac{\langle R_{\perp}\rangle}{c\sqrt{\pi}}\simeq3.0\times10^{[(m_{\rm uv}/26.66)^{-9.52}-1]}\,{\rm Myr}$ ($\simeq5.6$ Myr at $m_{\rm BB}^{\rm lim}=26.0$), we can only place an upper limit to the quasar lifetime in agreement with our $t_{\rm age}=1\rm\,Myr$ case. On the other hand, at a large quasar lifetime, the size of quasar transverse proximity zone exceeds that of the field of view of the assumed instrument.
As the quasar proximity zone size approaches the size of the field of view, only lower limit can be placed since there is no background galaxy sightline probing the region outside the transverse proximity zone. Note that in our mock survey field of view is limited by the size of our cosmological simulation ($146\rm\,cMpc$ on side, corresponding to light crossing time of $44\,\rm Myr$ from the centre to the edge of the box).
The same trend is expected for the field of view of Subaru/HSC,
giving a sensitivity to quasar lifetime below $t_{\rm age}<D_{\rm A}(z)\theta_{\rm FoV}/c\simeq60 (\theta_{\rm FoV}/45')\rm Myr$ at $z=4.4$ where $D_{\rm A}(z)$ is the angular diameter distance and $\theta$ is the angular radius of the field of view. In summary, the range of quasar lifetime that can be constrained by the photometric light-echo tomography is
\begin{equation}
    3.0\times10^{[(m_{\rm uv}/26.66)^{-9.52}-1]}\,{\rm Myr}<t_{\rm age}<60\left(\frac{\theta_{\rm FoV}}{45'}\right)\rm\,Myr,
\end{equation}
at $z=4.4$. The precision on $t_{\rm age}$ is set by the inter-sightline separation of background galaxies (equation \ref{eq:time_resol}).

\begin{figure}
	\includegraphics[width=\columnwidth]{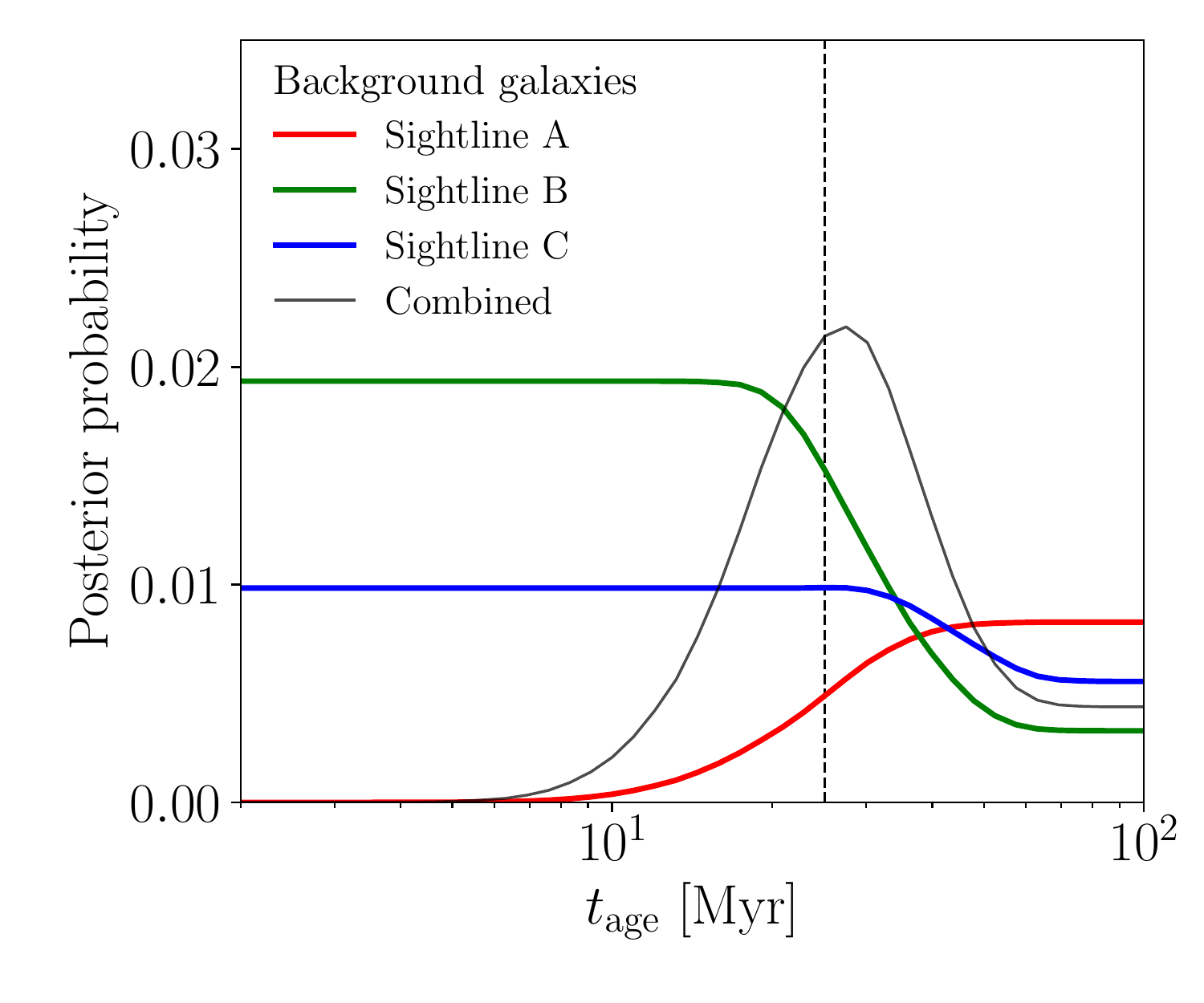}
	\vspace{-0.8cm}
\caption{Posterior probabilities of single background galaxies for NB tomography around a $z=4.4$ quasar with lifetime $t_{\rm age}=25.1\rm\,Myr$ (vertical dashed line) with the limiting BB magnitude of $m_{\rm BB}^{\rm lim}=26.0(5\sigma)$ and foreground NB magnitude of $m_{\rm BB}^{\rm lim}=27.4(3\sigma)$. The posterior from each background galaxy sightline A, B, C is indicated by the coloured lines, which corresponds to the sightlines labeled in Figure \ref{fig:2D_map}. The combined posterior from these three sightlines is indicated by thin black line.}
    \label{fig:single_posterior}
\end{figure}

Before discussing how the lifetime constraint depends on the survey depth, it is worthwhile to understand how the posterior probability is determined from the NB IGM tomography. The formation of the full posterior constraint on quasar lifetime from a collection of many individual background galaxy sightlines is illustrated in Figure \ref{fig:single_posterior}. The final posterior from multiple background galaxies is determined by the combination of upper and lower limits on the quasar lifetime from single background galaxies. If a background galaxy (e.g. sightline A) passes through a highly transmissive region of the IGM ($T_{\rm IGM}\gtrsim0.4$), then it sets a lower limit to the quasar lifetime according to the projected distance between the background galaxy and the foreground quasar, $t_{\rm age}>R_{\perp}/c$. On the other hand, if the background galaxy sightline (e.g. B or C) passes through an opaque region of the IGM ($T_{\rm IGM}\lesssim0.4$) close to the mean IGM transmission, it sets an upper limit $t_{\rm age}<R_{\perp}/c$. Collectively, when using multiple background sightlines, the final posterior peaks around the true value. Note that when the background sightline passes through a rare fictitious transmissive region due to the IGM density or photometric fluctuations, the single galaxy posterior sets a bound on the quasar lifetime at the wrong value. However, this will be automatically corrected by many other sightlines which effectively averages out the error coming from these sources of uncertainties, placing a final posterior at the right value.

\subsection{Impact of survey depths on lifetime constraint}\label{sec:depth}

\begin{figure*}
	\includegraphics[width=\columnwidth]{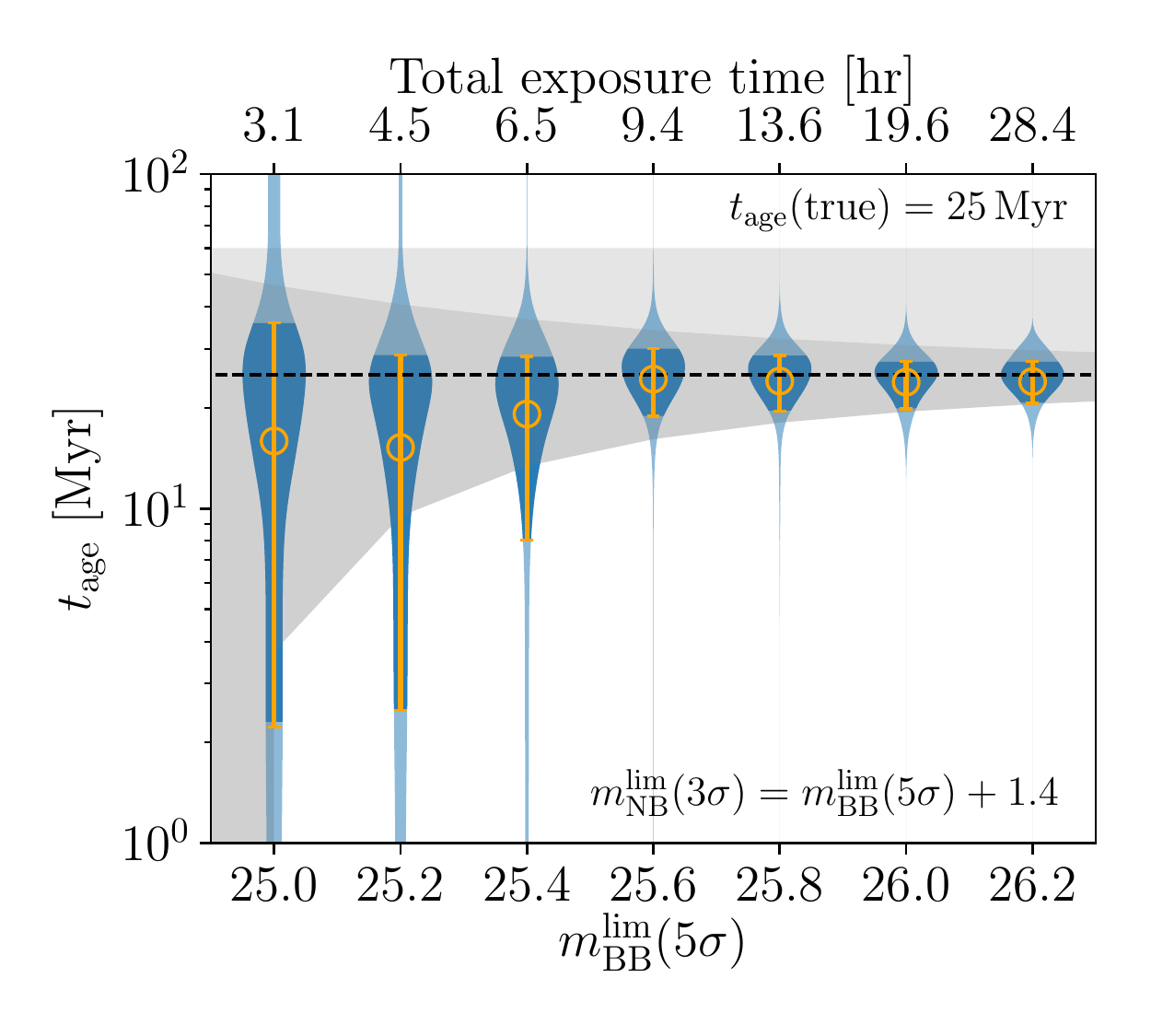}
	\includegraphics[width=\columnwidth]{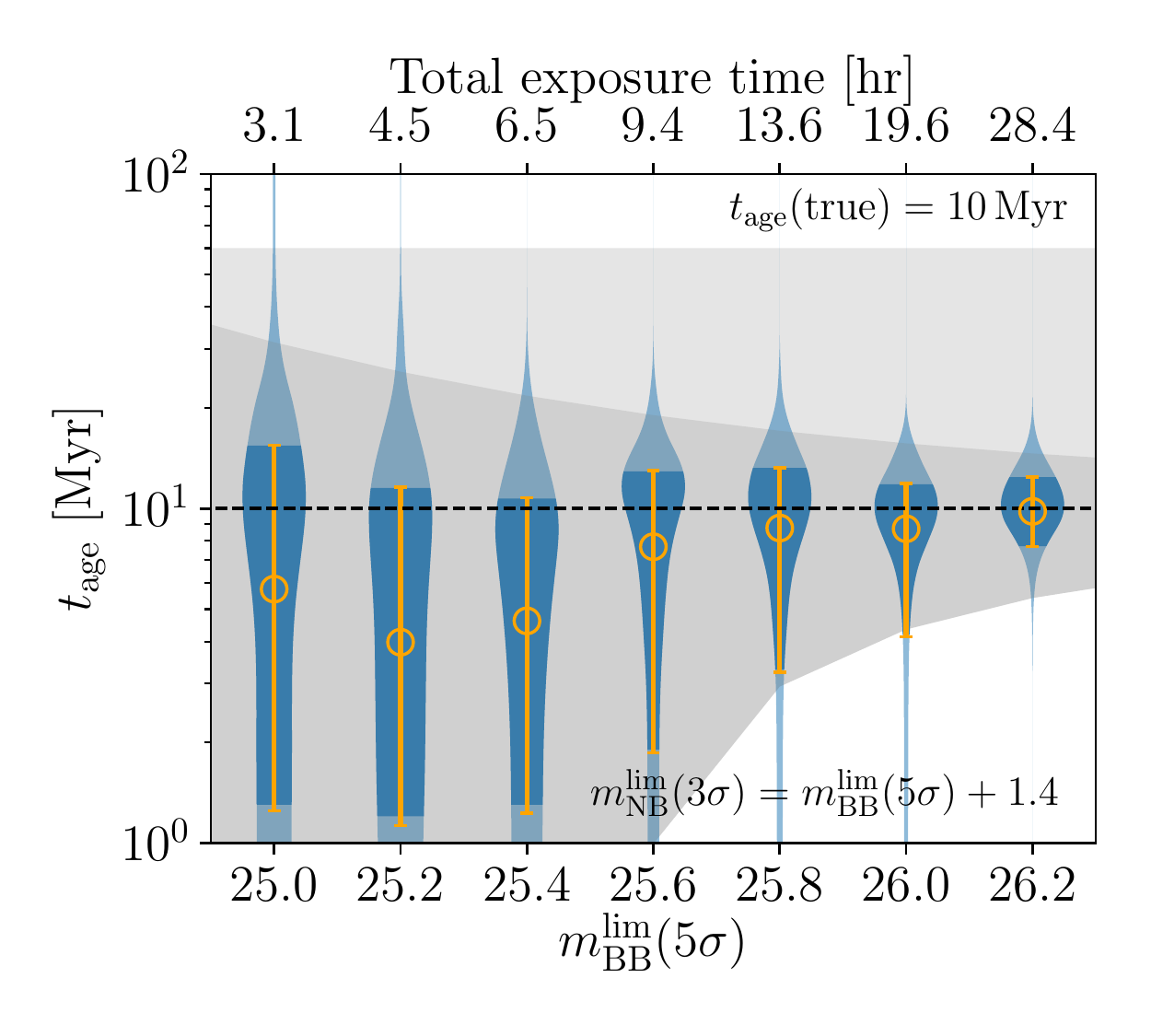}
	\vspace{-0.5cm}
\caption{The mean posteriors as a function of BB ($i2$) magnitude limits for two different quasar lifetime ($t_{\rm age}=25\,\rm Myr$ (left), $t_{\rm age}=10\,\rm Myr$ (right)) from NB tomography at $z=4.4$. The dashed lines indicate the true input value of the lifetime. The width of blue shaded regions indicate the probability at the corresponding $t_{\rm age}$ constraint. The errorbar and the dark blue shaded region indicate $68\%$ confidence interval. The dark gray shaded regions show the expected precision from the simple analytic estimate, and the light gray region indicate the upper bound due to the field-of-view. The total exposure time including NB656, NB718, $r2$, and $i2$ is shown in top x-axis. Note that the apparent bias to lower $t_{\rm age}$ value at shallower BB magnitude depth simply reflects the prior, indicating that there is no constraining power.}.
    \label{fig:violin_plot_mBB}
\end{figure*}

The BB magnitude depth is critical as it sets the number of background galaxies, determining the spatial and temporal resolution for the photometric light-echo tomographic experiment. Figure \ref{fig:violin_plot_mBB} shows the mean posteriors as a function of different BB magnitude depths for two different quasar lifetimes. The fiducial NB magnitude is set such that it gives the $3\sigma$ sensitivity to the mean IGM tranmission at a given BB magnitude depth. The 68 \% confidence interval on the quasar lifetime scales with the BB magnitude depth. This agrees with the expectation from our estimate in Section \ref{sec:background_sources},
\begin{equation}
    \Delta t_{\rm age}=\pm3.0\times10^{[(m_{\rm BB}/26.66)^{-9.52}-1]}\,{\rm Myr}.
\end{equation}
The envelop on the expected precision matches with the full calculation as illustrated by Figure~\ref{fig:violin_plot_mBB}. A shallower BB magnitude clearly limits our ability to constrain a shorter quasar lifetime. Below $m_{\rm BB}^{\rm lim}=25.6$ we lose the sensitivity to $t_{\rm age}<10\rm\,Myr$ systems as shown in the right panel of Figure~\ref{fig:violin_plot_mBB}. We can only provide an upper limit to the quasar lifetime,
reflecting the fact that we will be unable to detect a transmissive sightline due to the sparse
spatial sampling of the background LAEs. Thus, to constrain
the quasar lifetimes in the range of a few Myr to tens of Myr,
we require the BB limiting magnitude of
$m_{\rm BB}^{\rm lim}\gtrsim25.8(5\sigma)$. This determines the required NB magnitude to measure the Ly$\alpha$ forest transmission along the background sources, resulting $m_{\rm NB}^{\rm lim}\gtrsim27.2(3\sigma)$.

We convert the required depths into a total exposure time using the HSC exposure time
calculator (version 2.2)\footnote{\url{https://hscq.naoj.hawaii.edu/cgi-bin/HSC_ETC/hsc_etc.cgi}}. For a $1.0''$ seeing and transparency of $0.70$, Moon phase $7$, and Moon distance $90.0$ degree, and photometric aperture with $1.5''$ diameter,
we obtain
\begin{equation}
    m^{\rm lim}_{\rm NB656}(3\sigma)\simeq27.16+2.5\times\frac{1}{2}\log_{10}\left(\frac{t_{\rm exp}}{10\,\rm hr}\right),
\end{equation}
for the NB656 filter, and
\begin{equation}
    m^{\rm lim}_{i2}(5\sigma)\simeq25.78+2.5\times\frac{1}{2}\log_{10}\left(\frac{t_{\rm exp}}{0.4\,\rm hr}\right),
\end{equation}
for the $i2$ filter, which we use as a BB filter to measure the UV continua of background LAEs.

Assuming we apply the standard LAE selection at $z\simeq4.9$ using the NB718 filter \citep{Zhang2020,Ono2021}, we additionally require a minimal filter set of
of NB718 and $r2$ together with $i2$. Conservatively, we require NB718 depth to be equally deep as $i2$ to securely detect the
Ly$\alpha$ emission of the background galaxies and $r-i>0.8$ for dropout criterion;
\begin{equation}
    m^{\rm lim}_{\rm NB718}(5\sigma)\simeq25.78+2.5\times\frac{1}{2}\log_{10}\left(\frac{t_{\rm exp}}{1.8\,\rm hr}\right),
\end{equation}
for the NB656 filter, and
\begin{equation}
    m^{\rm lim}_{r2}(5\sigma)\simeq26.56+2.5\times\frac{1}{2}\log_{10}\left(\frac{t_{\rm exp}}{0.5\,\rm hr}\right),
\end{equation}
for the $r2$ filter.

The minimal requirement including NB718, $r2$, $i2$ to select the background LAEs at $z=4.9$ and NB656 to measure the Ly$\alpha$ forest transmission at $z=4.4$, the total exposure time required is $\gtrsim13\rm\,hrs$ for NB tomography per field. This is a modest cost compared to the spectroscopic tomography \citep{Schmidt2019},
making it a viable alternative to examine the ionizing lightcurve of a high-redshift quasar, i.e. the luminous growth history of the central SMBH.

\begin{figure*}
	\includegraphics[width=\columnwidth]{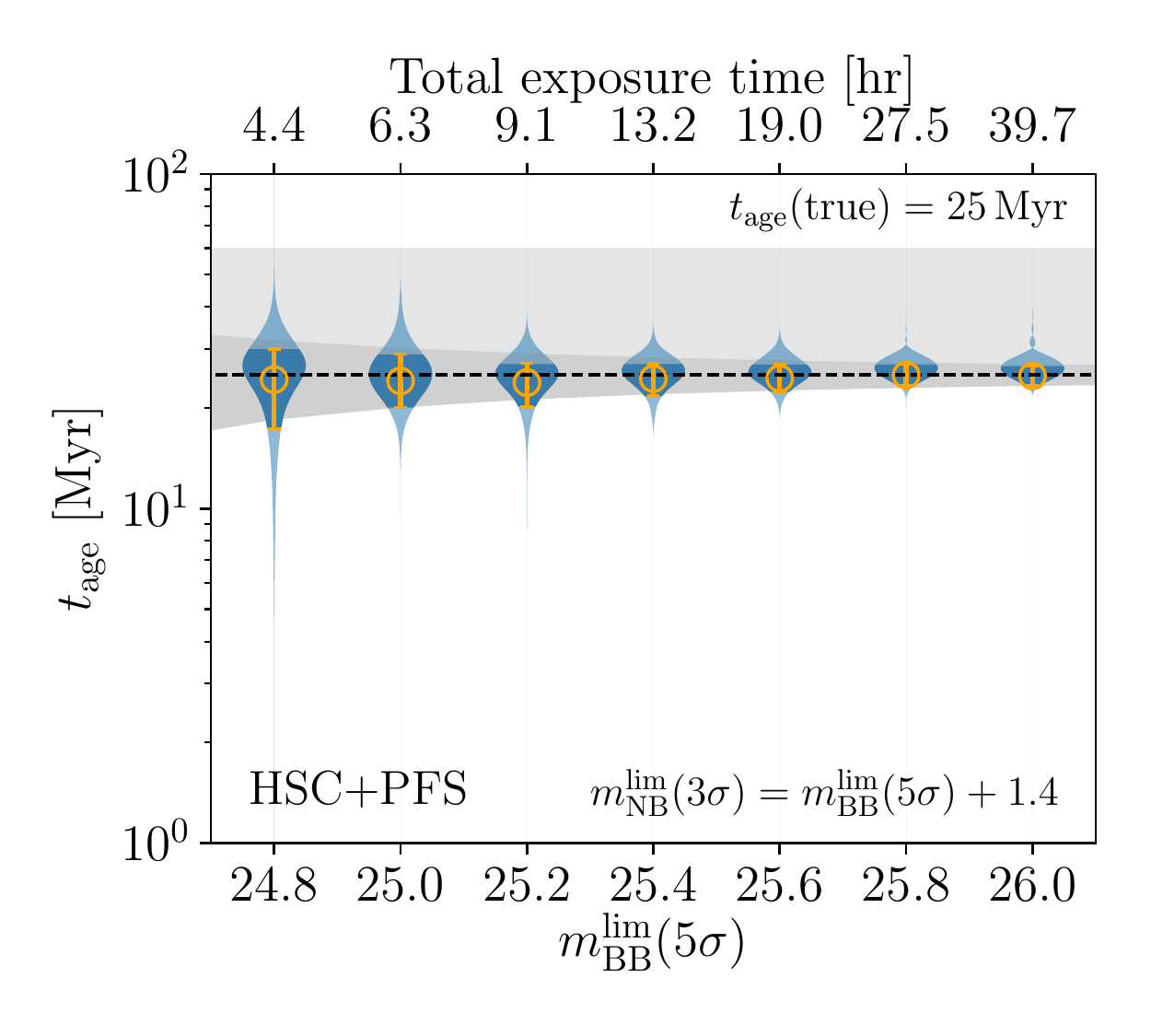}
	\includegraphics[width=\columnwidth]{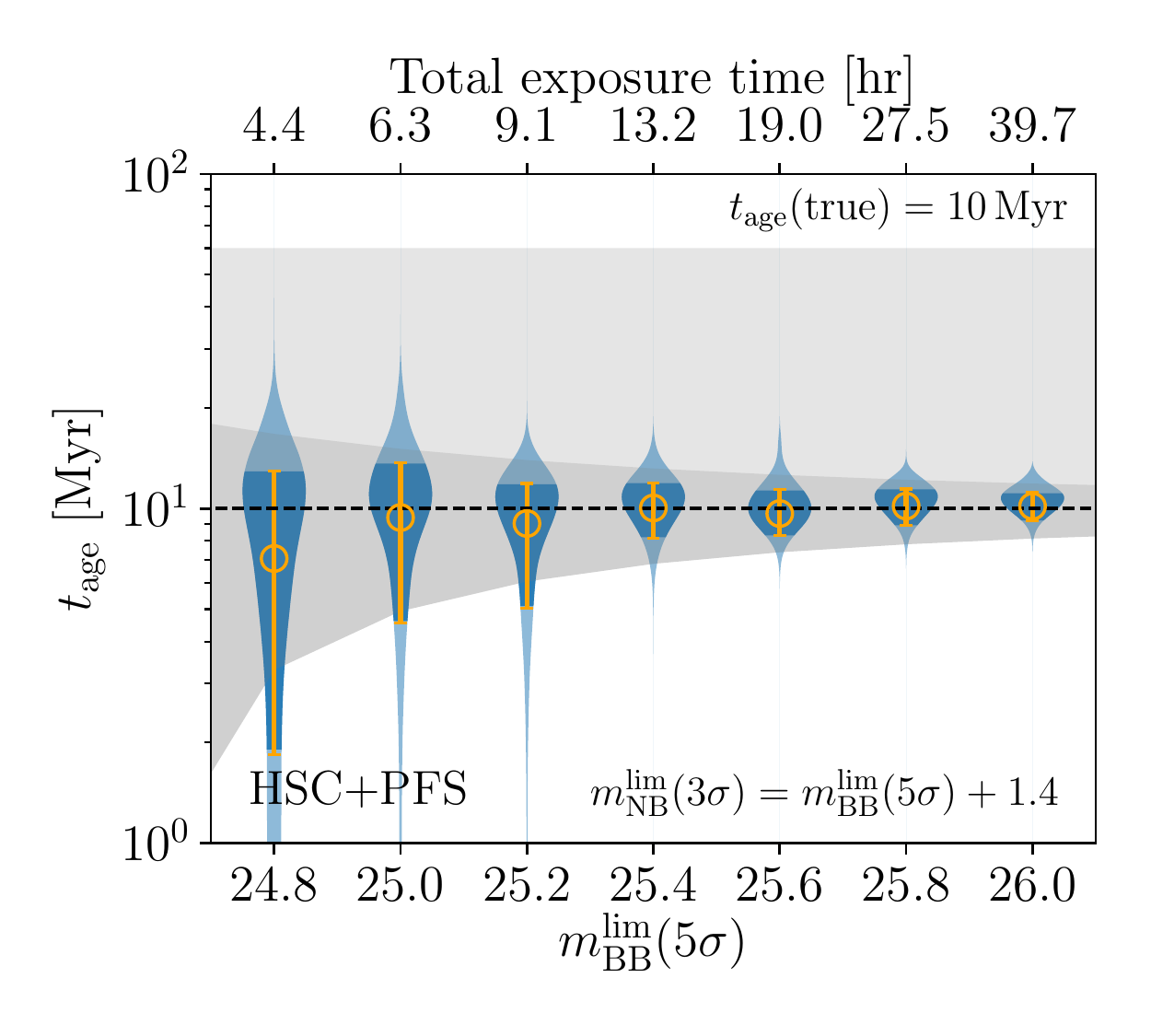}
	\vspace{-0.5cm}
\caption{Same as Figure~\ref{fig:violin_plot_mBB}, except that the number of background galaxies are computed assuming that spectroscopic galaxies will be identified by Subaru/PFS using Ly$\alpha$ emission. The total exposure time includes HSC $riz$ imaging to select $r-$-dropout and PFS spectroscopy to detect Ly$\alpha$ line from the dropout sample. The analytic estimate for the expected precision (dark gray shaded region) is computed using $\Delta t_{\rm age}=\langle R_\perp\rangle/(c\sqrt{\pi})=4.75\times10^{\left[(m_{\rm uv}/25.25)^{-14.05}-1\right]}\rm Myr$.}\label{fig:violin_plot_mBB_PFS}
\end{figure*}

\section{Discussion}\label{sec:discussion}

\subsection{Photometric or spectroscopic background sources}

We have examined the concept of NB tomography using phtometrically-identified LAEs as background sources. An alternative approach would be to use spectroscopic background galaxies, and measure the transmitted Ly$\alpha$ forest fluxes towards these sightlines using a NB filter. For example, this hybrid approach was employed by \citet{Mawatari2017} for the study of protoclusters at $z=3.1$ in the SSA22 field using publicly available catalogues of spectroscopic galaxies.
For quasar light-echo tomography, as we are likely targeting a new quasar field without a prior spectroscopic sample, in order to compare the two approaches, we need to estimate the observational costs required to assemble background spectroscopic sample.

We consider a case for Subaru/Prime Focus Spectrograph (PFS) scheduled to be operational in 2023. The spectroscopic redshift range suitable for NB tomography with NB656 is $z_1=4.50<z<z_2=5.30$ where the Ly$\alpha$ forest range of the background galaxies is covered by the NB656 filter. Confirming the galaxy redshifts via Lyman break feature is hard at these redshifts. The spectroscopic confirmation of dropout candidates thus relies on detecting Ly$\alpha$ emission lines. Applying the formalism discussed in Section \ref{sec:background_sources}, the expected background sightline surface density of background galaxies spectroscopically confirmed by Ly$\alpha$ emission lines is given by
\begin{equation}
    \Sigma_{\rm specz}=\int_{z_1}^{z_2}dz\left|\frac{dl}{dz}\right|(1+z)^3\int_{-\infty}^{\Muv^{\rm lim}}\frac{dn_{\rm\scriptscriptstyle LAE}}{d\Muv}d\Muv,
\end{equation}
where the Ly$\alpha$ equivalent width is $\rm REW>25$\AA~corresponding the value typically identified by a spectroscopic follow-up campaign.
This gives a higher spatial resolution of
\begin{equation}
    \langle R_\perp\rangle=\Sigma_{\rm specz}^{-1/2}\approx2.58\times10^{\left[(m_{\rm uv}/25.25)^{-14.05}-1\right]}\rm pMpc.
\end{equation}
This is approximately a factor of 2-3 times better than the original resolution using photometrically-identified background LAEs.
This increase in the spatial resolution comes from the increased line-of-sight volume of the spectrscopic background galaxies comapred to that of background LAEs. As a result, one can acheive a high spatial resolution of NB IGM tomography with bright spectroscopic background galaxies comparable to that using faint background LAEs. For example, the surface number density of spectroscopic background galaxies with $m_{\rm UV}<25.2$ is  $\Sigma_{\rm specz}\approx0.12\rm\,pMpc^2$, which is comparable with that obtained by a deep LAE sample with $m_{\rm UV}<26.0$ ($\Sigma_{\rm LAE}\approx0.11\rm\,pMpc^2$).

The Ly$\alpha$ luminosities of the spectroscopic galaxies with $\rm REW>25$\,\AA~and $m_{\rm UV}<25.2$ at redshift $4.5<z<5.3$ is $L_{\alpha}\gtrsim6.2\times10^{42}\rm\,erg\,s^{-1}$ corresponding to the observed flux of $\gtrsim2.4\times10^{-17}\rm \,erg\,s^{-1}\,cm^{-2}$. Given the expected $5\sigma$ line sensitivity of Subaru/PFS of $\simeq1.2\times10^{-17}\rm \,erg\,s^{-1}\,cm^{-2}$ for $6300-7500$\,\AA~range with $R\sim3000$ and 1 hour on-source exposure\footnote{\url{https://pfs.ipmu.jp/research/performance.html}} for the line width of $70\rm\,km\,s^{-1}$, assuming a typical Ly$\alpha$ line width of $240\rm\,km\,s^{-1}$, we expect to detect the Ly$\alpha$ line with $\sim3$ hours of exposure. The large $\sim1.25\,\rm deg^2$ field of view and high multiplexing with 2394 fibers of PFS should permit a single visit of single quasar field to be sufficient to secure spectroscopic sample by following up the dropout candidates in the field. This provides an efficient way to assemble a large sample of bright background galaxies suitable for NB IGM tomography.

Bright background galaxies also relax the requirement for the foreground NB depth. To achieve the same relative depth in NB656 compared to the BB (e.g. $m_{\rm BB}^{\rm lim}(5\sigma)=25.2$), we now only require $m_{\rm NB}^{\rm lim}(3\sigma)=m_{\rm BB}^{\rm lim}(5\sigma)+1.4=26.6$, corresponding to 3.5 hours exposure in NB656 instead of 15 hours for NB tomography using $m_{\rm BB}(5\sigma)<26.0$ background sources. Even taking into account the required cost for $riz$ imaging to identify $m_{\rm BB}<25.2$ $r$-dropouts\footnote{We assumed that $z$ band is used to measure the continuum level, and using the standard $r$-dropout selection \citep{Ono2018}. Using the same setup for the HSC ETC, we assumed $z$ band depth scales as
\begin{equation}
    m^{\rm lim}_{z}(5\sigma)\simeq25.35+2.5\times\frac{1}{2}\log_{10}\left(\frac{t_{\rm exp}}{1\,\rm hr}\right).
\end{equation}
} ($\sim3$ hrs), total HSC+PFS cost (9.1 hrs) is more than a factor of 2 cheaper compared to the HSC only (19.6 hrs) tomography.

Thus, the combination of Subaru/HSC+PFS provides an efficient approach for light-echo tomography. The expected constraining power of NB IGM tomography using spectroscopic background galaxies is shown in Figure\,\ref{fig:violin_plot_mBB_PFS}. It illustrates the significant reduction of total exposure time which includes all necessary HSC imaging and PFS spectroscopic follow-up to achieve the same constraining power as the NB IGM tomography only using background LAEs. We should contrast this with a pure spectroscopic IGM tomography where direct spectroscopic detections of the UV continua in background galaxies are required. In NB IGM tomography we only use PFS to confirm the redshifts of background galaxies. This drastically reduces the required observational cost. The high throughput of NB and BB imaging is used to measure the continuum and transmitted Ly$\alpha$ forest fluxes towards the spectroscopic background galaxies.

\subsection{Selection function and contamination}

In this paper, we assumed the zero contamination rate of background LAEs. However in realitiy there would be a finite contamination in the LAE sample because of the foreground interlopers such as low-$z$ $[\OII]$ or H$\alpha$ line emitters. The contamination rate of the LAE selection technique is estimated to be $\sim10-20\,\%$ \citep{Ouchi2008,Shibuya2018}. A low-$z$ interloper could introduce a fictitious Ly$\alpha$ forest transmission. However, as we require a detection in the UV continuum band for background sources, it additionally requires the Lyman break feature in the background galaxies. This may improve the purity of our background LAE sample. Furthermore, as we use multiple background LAEs to search for the quasar light echo signal, the impact of the contamination is expected to be minimal. Clearly the spectroscopic confirmation of the transmissive IGM sightlines along the background LAE candidates in the NB tomography is desirable.  We reserve the detail examination of the impact of the contamination in future work. Nonetheless, this should be considered as a show-stopper. NB tomography clearly provides an efficient method to map out a large region of the sky at once, which is difficult or observationally very expensive to achieve with traditional spectroscopic tomography.

\subsection{Emission geometry: position of foreground NB filter}

The anisotropic emission of ionizing photons from the quasar impacts the structure of the light echoes in the NB tomography. Our fiducial model in this paper assumes isotropic emission from the quasar, which maximises the region of influence of the ionizing light echoes on the IGM. The studies of the observed fraction of AGN ($F_{\rm obscured }\sim50\%$ e.g. \citealt{Lusso2013}) indicates that the opening angle of the quasar emission is statistically $\theta=2\arccos F_{\rm obsc}\approx120^\circ$. This means that some fraction of NB window may be obscured from the impact of quasar's ionizing radiation. Assuming that a quasar is located exactly at the centre of the filter and is directly pointing towards us, a simple geometric consideration finds the fraction of obscured segment of a line-of-sight passing through the NB filter is given by
\begin{equation}
    f_{\rm obsc}=\frac{2}{\tan\theta/2}\frac{R_{\perp}}{L_{\rm filter}},
\end{equation}
where $L_{\rm filter}$ is the comoving length of the NB filter. For NB656, the length corresponding to the FWHM is $L_{\rm filter}=56.6\rm\,cMpc$, meaning that $ f_{\rm obsc}\approx0.41(R_\perp/20\rm\,cMpc)$ for $\theta=120^\circ$. This will reduce the excess NB-integrated Ly$\alpha$ forest transmission as $f_{\rm obsc}$ fraction of a background sightline is obscured, given the net observed NB transmission $T_{\rm NB}$ to be $T_{\rm NB}=T_{\rm IGM}f_{\rm obsc}+T_{\rm IGM+QSO}(1-f_{\rm obsc})$ where $T_{\rm IGM+QSO}$ ($T_{\rm IGM}$) is the Ly$\alpha$ forest transmission of the IGM with (without) the impact of the quasar.

This issue can be circumvented by placing a foreground NB filter slightly foreground of the quasar redshift. This allows the NB filter to cover more fraction of the IGM impacted by the quasar ionizing radiation emitted towards the unobscured region with a minimal impact of the geometric $\propto r^{-2}$ dilution. For example, by probing the IGM with the NB filter just at the foreground of a quasar whose redshift matches the upper 50\% transmission wavelength of the NB filter,
we can reduce the impact of obscuration by a factor of 2, finding $ f_{\rm obsc}=\frac{1}{\tan\theta/2}\frac{R_{\perp}}{L_{\rm filter}}$. Indeed, in the serendipitous NB detection of the (diagonally) transverse proximity effect around a $z\simeq5.8$ quasar along the sightline of a bright background galaxy by \citet{Bosman2020}, the NB filter is located slightly in the foreground of the quasar.

The emission geometry introduces a source of degeneracy in the constraints on the quasar lightcurve. Note however that any positive detection of a transmissive sightline at an impact parameter $r_\perp$ immediately implies that the quasar was active at $\Delta t=r_\perp/c$ time in past.
The obscuration introduces complications for the interpretation of the null detection of sightlines with excess transmission, which can either be interpreted as quasar inactive phase or obscuration. This degeneracy can be lifted by obtaining deep spectroscopic follow-up of background galaxies. As spectra give multiple anchor points along a single lines-of-sight, it has a more constraining power on the emission geometry. This would break the degeneracy between emission geometry and radiative history of a quasar. Quantitative constraints combining both NB and spectroscopic tomographies need to be examined in future work.

\subsection{Past AGN activity of high-redshift quasars and galaxies}

The recent studies of quasar lifetime using the line-of-sight proximity effect indicate that the quasar lifetime is short on average $t_{\rm age}\sim10^6\rm\,yr$ \citep{Morey2021,Khrykin2021} with a fraction of the population having even shorter lifetime \citep{Eilers2017,Eilers2020,Eilers2021}. As noted in Section \ref{sec:intro}, this is shorter than the Salpeter timescale and poses a challenge to grow SMBHs at $z\gtrsim6$. The shortest lifetime that can be probed by the photometric light-echo tomography is limited by the mean sightline separation of background galaxies; for example, the sensitivity is $t_{\rm age}\simeq10\rm\,Myr$ for $m_{\rm BB}<25.5$ background sources. However, the real advantage of the photometric light-echo tomography is its ability to probe all past quasar activities in the last $\sim10^8\rm\,Myr$ limited by the field of view of the tomographic experiment. Each background sightline separated by $r_{\perp}$ from the quasar is sensitive to the light-echo produced at $t_{\rm age}\simeq r_\perp/c$. A detection of transmissive sightline in a photometric tomography experiment around a short lifetime quasar would indicate that the SMBH underwent another episode of luminous quasar activity in the past. This could potentially resolve the question of how $\sim10^9\rm\,M_\odot$ black holes can be present in the `young' quasars
uncovered in the recent work \citep{Eilers2017,Eilers2020,Eilers2021}, implying that photometric light-echo tomography around these objects is an exciting direction for future work.

While we consider targeting a (type I) quasar to probe the light curve and its past AGN activity in this paper, in principle we can generalise light-echo tomography around type II quasars and star-forming or quiescent galaxies at high redshifts. The black hole mass - stellar mass relation indicates that massive galaxies with stellar mass of $M_\ast>10^{10}\rm\,M_\odot$ host a supermassive black hole of mass $M_{\rm BH}>10^{9}\rm\,M_\odot$.
It has been argued that the Milky Way went through a period of luminous AGN activity in the past \citep{Bland-Hawthorn2019}. Some local galaxies show evidence for past AGN activity through the light echo emission from the CGM \citep{Keel2012,Keel2015,Keel2017}, and in addition, observed high ionization absorption lines in the CGM of massive galaxies may require flickering AGN activity of the central SMBH
\citep{Oppenheimer2018}. If quasar feedback is responsible for quenching high-redshift galaxies, then these objects might show evidence for AGN activity in the past. Light-echo tomography can probe the AGN ionising radiation emitted in all directions in past several tens of Myr. One could apply the proposed light-echo tomography in a well-studied extragalatic field to search for enhanced Ly$\alpha$ forest transmission as a sign of past AGN activity on the time baseline of $\sim10^8\rm\,Myr$ in past.

\section{Conclusions}\label{sec:conclusion}

In this paper we examined the capability of photometric IGM tomography to map the coherent spatial fluctuations of Ly$\alpha$ forest on the scale of $\sim10h^{-1}\rm cMpc$. We applied this technique to study the impact of quasar light echoes on the IGM to constrain the luminous growth history of individual SMBHs in $\sim\rm Myr$ timescale. This photometric technique uses a pair of NB filters carefully selected in order to measure the transmitted Ly$\alpha$ forest flux with the foreground NB filter along the background LAEs. The technique provides an economical observational strategy to create a two-dimensional map of the Ly$\alpha$ forest transmission, taking advantage of  the high throughput and wide field of view of imaging compared to spectroscopy to detect the faint transmitted Ly$\alpha$ forest flux over a large area of sky.
We examined the observational requirements of this photometric IGM tomography in detail. For double NB tomography, we require background LAEs with bright enough UV continua such that the Ly$\alpha$ forest flux decrements can be measured as the ratio of the foreground NB and the BB flux. Using an observationally-calibrated model for the LAE number density, we find a photometric NB survey with UV continuum depth of $m_{\rm UV}\sim25-26$ can achieve the mean inter-sightline separation of $\sim1-10\rm\,pMpc$ at $z\sim2-6$, which sets the spatial resolution of the tomographic map. We summarised the result in terms of the analytic fit to the expected mean inter-sightline separation of background LAEs at $z\simeq2.5-5.7$ (Table~\ref{tab:fit}). When applied to map quasar light echoes, this translates into a temporal resolution of $\sim3-30\rm,Myr$ for the quasar lightcurve. Thus, an NB imaging survey with NB and BB depths of $\sim25-26$ is sufficient to examine and search for the quasar light echo signal using photometric IGM tomography.

Besides the standard NB and BB filter combinations required to identify background LAEs, the foreground NB depth is critical to detect the transmitted Ly$\alpha$ forest flux along the background LAEs. As a rule-of-thumb, we find that the required limiting magnitude of the foreground NB filter is set by
\begin{equation}
m_{\rm NB}^{\rm lim}\approx m_{\rm BB}^{\rm lim}-2.5\log_{10}T_{\rm IGM},
\end{equation}
where $T_{\rm IGM}$ is the mean IGM transmission at the redshift of interest. We examined the required foreground NB depth more carefully using mock observations based on cosmological simulations. We find that a modest signal-to-noise of $\sim3$ at the limiting NB depth is sufficient to recover the 2D Ly$\alpha$ forest transmission map of the quasar light echo at $z=4.4$.
The other sources of noise in light-echo tomography such as the UV continuum error and IGM fluctuations are sub-dominant compared to the photometric error; they pose no obstacle to our
capability to detect the light-echo signal in the NB tomography.

We then introduced a fully Bayesian framework to infer the quasar lifetime from a NB tomographic survey. The framework captures the full information of the survey since it acts directly on the data (i.e. photometric dataset of background LAEs) instead of the processed data such as the azimuthally-averged Ly$\alpha$ forest transmission profile around a quasar or reconstructed 2D tomographic map. The framework is designed to simultaneously constrain both the quasar lifetime and the intrinsic SED of background LAEs and to rigorously propagate the uncertainties from photometric noise, IGM fluctuations, and intrinsic SED shape into the final measurement. This is possible because we can forward model the expected quasar light-echo signal using cosmological simulations including realistic noise. The framework is generalisable to include multiple BB filters to reduce, for examine, the error from the UV continuum slope uncertainty if necessary.

Applying the Bayesian framework to mock observations, we find that a NB tomographic survey of background NB and BB depth of $\simeq26.0(5\sigma)$ and foreground NB depth of $\simeq27.4(3\sigma)$ can constrain the quasar lifetime to $\sim20\,\%$ accuracy at $z=4.4$ for the range of $t_{\rm age}\sim10-50\rm\,Myr$. The precision of the lifetime constraint depends the number of background LAEs with a continuum detection. Thus it scales with the BB depth. Including all the required NB and BB filters for a NB tomographic survey, we find that for Subaru/HSC a total exposure time of $\sim13\rm\,hrs$ per field is required to tomographically map the quasar light-echoes.
While this is relatively large investment of telescope time, it is modest compared to spectroscopic tomography which would require a significantly more expensive deep spectroscopic follow-up campaign end pre-imaging. Future synergy with Subaru/PFS will significantly increase the number of background galaxies suitable for NB tomography, improving the accuracy and reducing the observational cost of the tomographic survey. We emphasise that light-echo tomography is currently the only viable technique to probe the quasar lightcurve over the timescale comparable to the Salpeter timescale, providing an indispensable tool to measure the luminous growth history of a SMBH at high redshift.

\section*{Acknowledgements}

We thank Suk Sien Tie for carefully reading the manuscript and Zarija Luki\'c for making the NyX simulation available for us. We would also like to thank the members of the ENIGMA group at UCSB and Leiden for constructive comments on an early version of this manuscript.
\section*{Data Availability}

The data and code used in this paper is available upon reasonable request to the authors.



\bibliographystyle{mnras}
\bibliography{reference} 




\onecolumn
\appendix

\section{one narrow-band + two (or multi-) broad-band survey strategy}\label{app:multi-band}

When we have two or more BB flux measurment of background galaxies, it is straightforward to generalise the Bayesian inference framework to simultaneously constrain the continuum slopes $\beta$ (i.e intrinsic SED shape) of background galaxies and the quasar lifetime $t_{\rm age}$. In this case, we will have multiple BB fluxes (e.g. $i$- and $z$-bands),
\begin{equation}
    f_{\rm BB1}^{\rm obs}=f_{\rm BB1}+\delta f_{\rm BB1},~~~~~
    f_{\rm BB1}=\frac{\int f_\nu  T_{\rm BB1}(\nu)d\nu}{\int T_{\rm BB1}(\nu)d\nu},
\end{equation}
and
\begin{equation}
    f_{\rm BB2}^{\rm obs}=f_{\rm BB2}+\delta f_{\rm BB2},~~~~~
    f_{\rm BB2}=\frac{\int f_\nu  T_{\rm BB2}(\nu)d\nu}{\int T_{\rm BB2}(\nu)d\nu}.
\end{equation}
Then the probability of observing NB and two BB fluxes for a $i$-th background galaxy is written as
\begin{align}
&P_i(f_{\rm NB}^{\rm obs},f_{\rm BB1}^{\rm obs},f_{\rm BB2}^{\rm obs}|t_{\rm age},\boldsymbol{r}_\perp,f_{\rm 1500},\beta,\sigma_{\rm NB},\sigma_{\rm BB1},\sigma_{\rm BB2})= \nonumber \\
&~~\mathcal{N}(f_{\rm BB1}^{\rm obs}|f_{\rm BB1},\sigma_{\rm BB1})\times\mathcal{N}(f_{\rm BB2}^{\rm obs}| f_{\rm BB2},\sigma_{\rm BB2})\times\int \mathcal{N}(f_{\rm NB}^{\rm obs}|f_{\rm NB},\sigma_{\rm NB})P(T_{\rm IGM}|\boldsymbol{r}_\perp,t_{\rm age})dT_{\rm IGM}.
\end{align}
The likelihood of observing a set of fluxes from $N$ background galaxies is then given by
\begin{align}
\mathcal{L}(\{f_{{\rm NB}}^{\rm obs}\, , f_{{\rm BB1}}^{\rm obs},f_{{\rm BB2}}^{\rm obs}\}_{i=1,\dots,N}|t_{\rm age},\{\boldsymbol{r}_\perp,f_{{\rm 1500}},\beta\}_{i=1,\dots,N},\sigma_{\rm NB},\sigma_{\rm BB}) \nonumber \\
~~~~~=\prod_{i=1}^N P_i(f_{{\rm NB}}^{\rm obs}\,, f_{{\rm BB1}}^{\rm obs},f_{{\rm BB2}}^{\rm obs}|t_{\rm age},\boldsymbol{r}_\perp,f_{{\rm 1500}},\beta,\sigma_{\rm NB},\sigma_{\rm BB1},\sigma_{\rm BB2})
\end{align}
The posterior of the quasar lifetime is therefore given by
\begin{align}
&P(t_{\rm age}|\{f_{{\rm NB}}^{\rm obs}\, , f_{{\rm BB1}}^{\rm obs}\, , f_{{\rm BB2}}^{\rm obs}, \boldsymbol{r}_\perp\}_{i=1,\dots,N},\sigma_{\rm NB},\sigma_{\rm BB1},\sigma_{\rm BB2})\propto  \nonumber \\
&~~~~~~\prod_{i=1}^N \int_0^\infty df_{1500} P(f_{1500})\int_\infty^\infty d\beta P(\beta) P_i(f_{{\rm NB}}^{\rm obs}\,, f_{{\rm BB1}}^{\rm obs}\,,f_{{\rm BB2}}^{\rm obs}|t_{\rm age},\boldsymbol{r}_\perp,f_{{\rm 1500}},\beta,\sigma_{\rm NB},\sigma_{\rm BB1},\sigma_{\rm BB2}).
\end{align}
We can generalise this Bayesian framework for multi-BB photometry and more general intrinsic SED model. This replaces the power-law spectrum parameterised by $f_{1500}$ and $\beta$ with those from the SED library of a stellar population synthesis model and we can use e.g. stellar age, metallicity, and star formation rate as an alternative parametrisation.


\bsp	
\label{lastpage}
\end{document}